\pdfoutput=1
\documentclass[times]{fldauth}

\usepackage{moreverb}

\usepackage[colorlinks,bookmarksopen,bookmarksnumbered,citecolor=red,urlcolor=red]{hyperref}
\usepackage{bm}
\usepackage{stmaryrd}
\usepackage{doi}

\usepackage{listings}
\definecolor{lbcolor}{rgb}{0.93,0.93,0.93}
\newcommand\BibTeX{{\rmfamily B\kern-.05em \textsc{i\kern-.025em b}\kern-.08em
T\kern-.1667em\lower.7ex\hbox{E}\kern-.125emX}}

\def\volumeyear{2010}

\begin{document}

\runningheads{B.~KRANK et al.}{Wall modeling via function enrichment within a high-order DG method}

\title{Wall modeling via function enrichment within a high-order DG method for RANS simulations of incompressible flow}

\author{Benjamin~Krank, Martin~Kronbichler and Wolfgang~A.~Wall\corrauth}

\address{Institute for Computational Mechanics, Technical University of Munich, Boltzmannstr. 15, 85748 Garching, Germany}

\corraddr{Institute for Computational Mechanics, Technical University of Munich, Boltzmannstr. 15, 85748 Garching, Germany. Tel.: +49 89 28915300; fax: +49 89 28915301; e-mail: \{krank,kronbichler,wall\}@lnm.mw.tum.de}

\begin{abstract}
We present a novel approach to wall modeling for RANS within the discontinuous Galerkin method. Wall functions are not used to prescribe boundary conditions as usual but they are built into the function space of the numerical method as a local enrichment, in addition to the standard polynomial component. The Galerkin method then automatically finds the optimal solution among all shape functions available. This idea is fully consistent and gives the wall model vast flexibility in separated boundary layers or high adverse pressure gradients. The wall model is implemented in a high-order discontinuous Galerkin solver for incompressible flow complemented by the Spalart--Allmaras closure model. As benchmark examples we present turbulent channel flow starting from $Re_{\tau}=180$ and up to $Re_{\tau}=100{,}000$ as well as flow past periodic hills at Reynolds numbers based on the hill height of $Re_H=10{,}595$ and $Re_{H}=19{,}000$.
\end{abstract}

\keywords{Wall modeling; XFEM; Wall functions; Discontinuous Galerkin; RANS; Spalart--Allmaras}

\maketitle

\vspace{-6pt}

\section{Introduction}

Turbulent wall-attached flows exhibit a sharp velocity gradient at no-slip boundaries. This gradient commonly requires the first off-wall point to be located within the viscous sublayer at $y_1^+\sim 1$ for Reynolds-averaged Navier--Stokes (RANS) simulations. Wall functions are therefore frequently used to economize computer time and storage requirements by placing the first node in the logarithmic region and specifying appropriate boundary conditions (see, e.g., the reviews in~\cite{Launder74,Durbin09}).

This approach comes along with two major challenges: (i) wall functions often fail to produce accurate results in strong non-equilibrium boundary layers and (ii) it may be difficult to define consistent boundary conditions for turbulence quantities at general off-wall locations, especially inside the buffer layer between $5< y^+ < 30$~\cite{Durbin09,Kalitzin05}. Special wall functions have been designed in order to enhance performance in non-equilibrium flow settings, e.g. in~\cite{Vieser02,Knopp06,Popovac07}. Suitable boundary conditions for turbulence quantities have been discussed, for example, in~\cite{Kalitzin05}.

In the present paper, we do not apply wall functions to prescribe boundary data. Instead, we construct a problem-tailored spatial discretization, which is capable of resolving sharp boundary layer gradients with coarse grids ($y_1^+ \sim 1{,}000$), while offering a much higher degree of flexibility in non-equilibrium conditions than wall functions can provide. This is achieved by consistently enriching the polynomial function space of the Galerkin method near a no-slip boundary with a wall function. The solution is here composed of the standard polynomial plus an enrichment, which in our work is constructed using Spalding's law-of-the-wall. The method then automatically ``chooses'' the most appropriate solution among the polynomial and the enrichment space. The resulting algorithm allows representation of general boundary layers including non-equilibrium conditions (cf. (i)) and resolves the whole velocity profile down to the no-slip boundary such that artificial boundary conditions of turbulence quantities are not required (cf. (ii)).

A framework for the development of such problem-tailored numerical methods was first proposed by Melenk and Babu\v{s}ka~\cite{Melenk96} with their partition-of-unity method (PUM). Belytschko and Black~\cite{Belytschko99} have subsequently suggested a formalism for enrichment of crack tip elements in solid mechanics within the continuous Galerkin method (standard FEM). The general version of the latter approach is denoted extended finite element method (XFEM) and has been employed in a number of applications; the interested reader is referred to the review article~\cite{Fries10} and previous studies in the field of high-gradient enrichments, e.g. for the convection-diffusion equation~\cite{Turner10} or Burger's equation~\cite{Chen14}. The so-called discontinuous enrichment method (DEM) by Farhat and co-workers~\cite{Farhat01} with application to high-gradient enrichments for the convection-diffusion equation, e.g. in ~\cite{Kalashnikova09}, represents another framework for constructing problem-tailored numerical schemes. A major distinction between these two methods is that XFEM is typically (but not necessarily) based on continuous functions both for the polynomial and the enrichment component, whereas in DEM the polynomial component is continuous and the enrichment is discontinuous at cell interfaces allowing for static condensation of the enrichment via Lagrange multipliers.

The present paper builds upon the boundary layer enrichment within the XFEM framework as recently proposed by Krank and Wall~\cite{Krank16} for the incompressible Navier--Stokes equations. This approach is developed further in the present work to RANS modeling in conjunction with high-order discontinuous Galerkin (DG) discretizations. Hence, both the standard and the enrichment component are discontinuous and the elements are coupled via appropriate fluxes as usual in DG. We prefer XFEM over DEM in the present application since it may be considered more efficient within the particular matrix-free implementation~\cite{Krank16b} used in this paper while the additional degrees of freedom only amount to a few percent of the overall number of degrees of freedom. The DG method has recently attained increasing popularity for computation of turbulent flow (see, e.g.,~\cite{Krank16b} and references therein) and additionally offers favorable characteristics regarding function enrichment. The Spalart--Allmaras model~\cite{Spalart94} is selected as RANS closure due to its simplicity and yet good performance as well as positive experiences within the DG context~\cite{Moro11,Burgess12,Crivellini13,Darmofal13,ZhenHua16}.

The outline of this article is as follows. Section~\ref{sec:ins} introduces the governing equations under consideration in this paper. Section~\ref{sec:wm} subsequently presents wall modeling via function enrichment within DG discretizations. The numerical method including time stepping, variational form, matrix formulation, and implementation of the enrichment is discussed in Section~\ref{sec:num}. Numerical benchmark examples consisting of plane turbulent channel flow and flow past periodic constrictions are presented in Section~\ref{sec:examples}, and concluding remarks close the article in Section~\ref{sec:con}.

\section{Incompressible Navier--Stokes Equations and the Spalart--Allmaras model}
\label{sec:ins}
In this work we consider the incompressible Navier--Stokes equations in conservative form with the Spalart--Allmaras~\cite{Spalart94} one-equation closure model:
\begin{alignat}{5}
&\nabla \cdot \bm{u} &&= 0  &&\text{ \hspace{0.2cm} in } \Omega \times [0,\mathcal{T}] \label{eq:conti} \\
\frac{\partial \bm{u}}{\partial t} + &\nabla \cdot \left(\bm{\mathcal{F}}^c(\bm{u}) + p \bm{I} - \bm{\mathcal{F}}^{\nu}(\bm{u})\right) &&= \bm{f} &&\text{ \hspace{0.2cm} in } \Omega \times [0,\mathcal{T}]\\
\frac{\partial \tilde{\nu}}{\partial t} + &\nabla \cdot \left(\tilde{\bm{\mathcal{F}}}^c(\bm{u},\tilde{\nu}) - \tilde{\bm{\mathcal{F}}}^{\tilde{\nu}}(\tilde{\nu})\right) &&= Q(\bm{u},\tilde{\nu})  &&\text{ \hspace{0.2cm} in } \Omega \times [0,\mathcal{T}].
\label{eq:mom}
\end{alignat}
Herein, $\bm{u}$ is the velocity, $p$ the kinematic pressure, $\tilde{\nu}$ the eddy-viscosity-like primary variable of the Spalart--Allmaras model, $\bm{f}$ the body-force vector, $\Omega$ the domain size, and $\mathcal{T}$ the simulation time.
The convective and viscous fluxes of the momentum equation are defined as
\[
\bm{\mathcal{F}}^c(\bm{u}) = \bm{u} \otimes \bm{u} \hspace{1cm} 
\bm{\mathcal{F}}^{\nu}(\bm{u}) = 2 (\nu + \nu_t) \bm{\epsilon}(\bm{u})
\]
with $\bm{\epsilon}(\bm{u}) = 1/2(\nabla \bm{u} + (\nabla \bm{u})^T)$, the kinematic viscosity $\nu$, and the eddy viscosity $\nu_t = \tilde{\nu} f_{v1}$ where $f_{v1}$ is a turbulence model quantity.
The convective as well as diffusive fluxes of the Spalart--Allmaras equation are given as
\[
\tilde{\bm{\mathcal{F}}}^c(\bm{u},\tilde{\nu}) = \bm{u} \tilde{\nu} \hspace{1cm} 
\tilde{\bm{\mathcal{F}}}^{\tilde{\nu}}(\tilde{\nu}) = \frac{\nu + \tilde{\nu}}{c_{b3}} \nabla \tilde{\nu}
\]
where $c_{b3}$ is a model constant. The formulation of the source term $Q(\bm{u},\tilde{\nu})$ employed in this paper including all model constants is described in Appendix~\ref{sec:sa_sourceterm}.

The initial conditions are specified at $t=0$ as
\begin{alignat}{3}
\bm{u}(t=0) &= \bm{u}_{0} &&\text{ \hspace{0.2cm} in } \Omega \text{ \hspace{0.2cm} and }\\
\tilde{\nu}(t=0) &= \tilde{\nu}_0 &&\text{ \hspace{0.2cm} in } \Omega.
\end{alignat}
This work considers periodic as well as no-slip Dirichlet boundary conditions on solid walls $\partial \Omega^D = \partial \Omega$ that close the problem with
\begin{alignat}{3}
\bm{u} &= \bm{g}_{\bm{u}} = \bm{0} &&\text{ \hspace{0.2cm} on } \partial \Omega^D \text{ \hspace{0.2cm} and }\\
\tilde{\nu} &= g_{\tilde{\nu}} = 0 &&\text{ \hspace{0.2cm} on } \partial \Omega^D.
\end{alignat}

\section{Wall modeling via function enrichment}
\label{sec:wm}
The sharp spatial velocity gradient present in turbulent boundary layers leads to resolution requirements of $y_1^+\sim 1$ for RANS simulations using standard low-order schemes. The aim of this work is to construct a problem-tailored numerical method that is capable of resolving high velocity gradients with very coarse meshes ($y_1^+\sim 1000$) while preserving flexibility in non-equilibrium boundary layers. This is achieved by inserting a wall function in the function space of the Galerkin method in addition to the standard polynomial component. It is inherent to all Galerkin methods that the user chooses the shape functions, and the method automatically tries to find an optimal solution using these shape functions. The general framework for such a function enrichment is presented in the following subsection (Section~\ref{sec:enrichment}), and the particular wall function used is discussed in Subsection~\ref{sec:spalding}. The reasoning largely follows Krank and Wall~\cite{Krank16} and introduces several simplifications possible in the framework of DG.

As a basis for the following discussion we tessellate the $d$-dimensional computational domain $\Omega_h \subset \mathbb{R}^d$ into $N_e$ non-overlapping quadrilateral or hexahedral finite elements $\Omega_h = \bigcup_{e=1}^{N_e} \Omega_e$ in $d=2$ or $d=3$, respectively. The subscript $(\cdot)_h$ used here indicates the identification of the respective variable with a characteristic element length $h$. The standard polynomial solution spaces are of the form
\begin{equation*}
\mathcal{V}_h^p = \{p_h \in L^2 : p_h |_{\Omega_e} \in P_k(\Omega_e), \forall e \in \Omega_h \},
\end{equation*}
here for example for the pressure $p_h$, where $P_k$  are all polynomials up to the tensor degree $k$. We choose nodal shape functions based on Gauss--Lobatto nodes throughout this paper due to reasons of efficiency~\cite{Krank16b}, while modal shape functions would be equally suitable for all DG spaces.

\begin{figure}[htb]
\center 
\includegraphics[trim= 9mm 3mm 9mm 0mm,clip,scale=0.7]{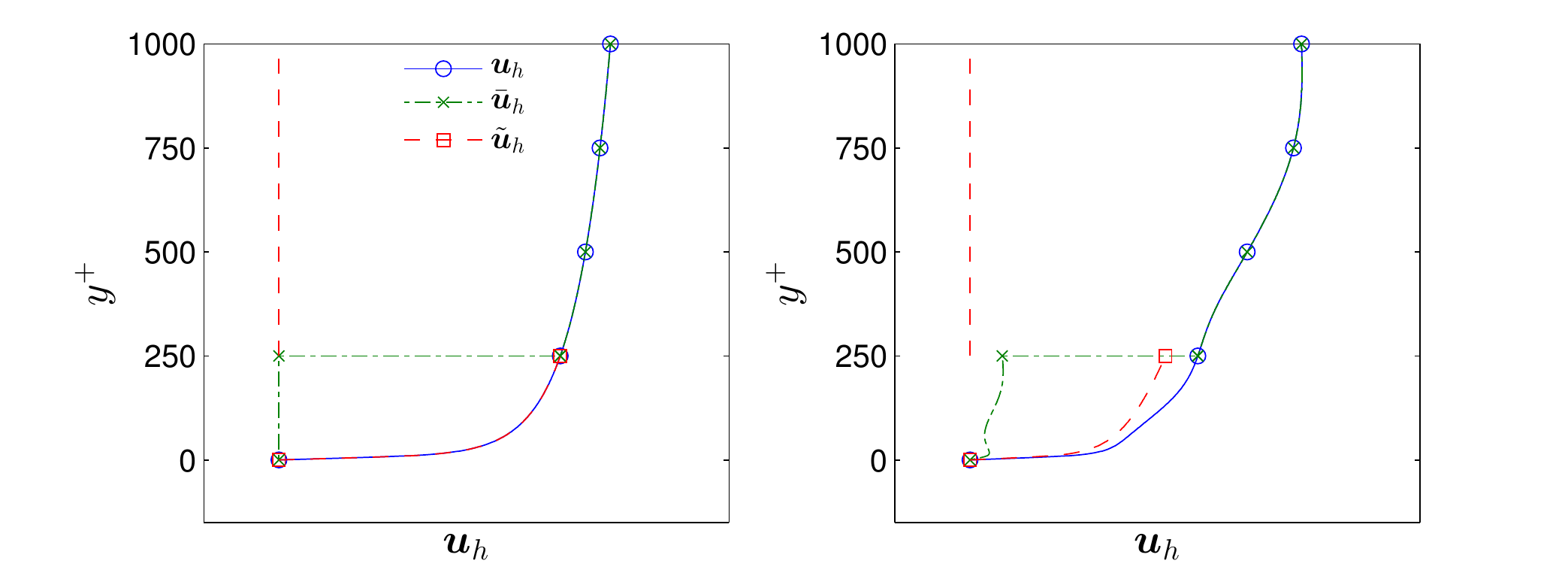}
\begin{picture}(100,0)
\put(132,84){\vector(-1,-1){23}}
\put(148,90){\vector(4,-1){20}}
\put(108,90){\footnotesize \parbox{1.5cm}{polynomial of degree $k$}}
\put(170,56){\vector(-4,1){20}}
\put(170,56){\footnotesize \parbox{2.1cm}{linearly\\weighted\\Spalding's law}}
\thicklines
\put(-101,40){\line(1,0){131}}
\put(85,40){\line(1,0){131}}
\multiput(-91, 40)(10, 0){13}{\line(-1,-1){10}}
\multiput(95, 40)(10, 0){13}{\line(-1,-1){10}}
\end{picture}
\caption{Composition of the velocity $\bm{u}_h$ in a boundary layer consisting of the polynomial and the enrichment component $\bar{\bm{u}}_h$ and $\tilde{\bm{u}}_h$ for an ideal boundary layer given as Spalding's law (left) and a non-equilibrium boundary layer in a separated flow or with high pressure gradient (right). Symbols indicate cell boundaries, i.e. one cell near the wall is enriched.}
\label{fig:decomp}
\end{figure}
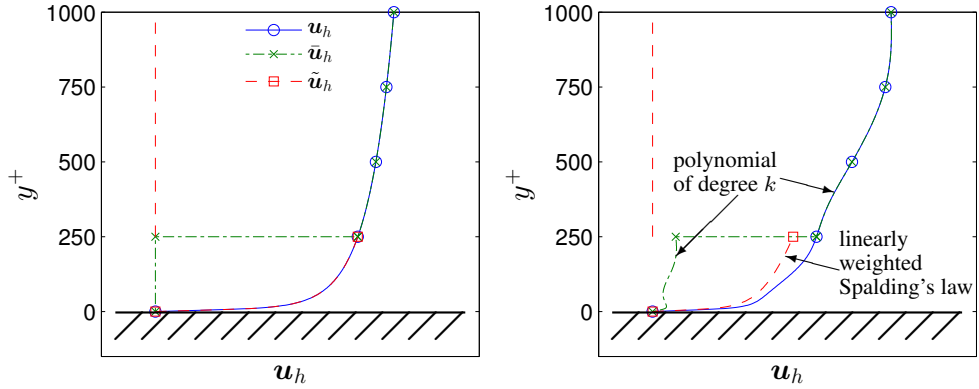

\subsection{Enriching the solution space}
\label{sec:enrichment}
The key idea of the present wall modeling approach is that the function space for the velocity $\bm{u}_h$ consists of two parts, a polynomial component $\bar{\bm{u}}_h(\bm{x},t)\in\mathcal{V}_h^{\bar{\bm{u}}}=(\mathcal{V}_h^p)^d$ plus an enrichment component $\tilde{\bm{u}}_h(\bm{x},t)$, yielding
\begin{equation}
\bm{u}_h(\bm{x},t)=\bar{\bm{u}}_h(\bm{x},t)+\tilde{\bm{u}}_h(\bm{x},t)
\label{eq:space}
\end{equation}
with the spatial location vector $\bm{x}$ and assuming the direct sum decomposition of the underlying discrete solution spaces $\mathcal{V}_h^{\bm{u}}=\mathcal{V}_h^{\bar{\bm{u}}} \oplus \mathcal{V}_h^{\tilde{\bm{u}}}$.
The polynomial component $\bar{\bm{u}}_h(\bm{x},t)$ may in each element be represented by a standard FE expansion of degree~$k$:
\begin{equation}
\bar{\bm{u}}_h(\bm{x},t)=\sum_{B \in N^{k}} N_B^{k}(\bm{x}) \bar{\bm{u}}_B(t)
\label{eq:std_fe}
\end{equation}
with shape functions $N_B^{k}$ and degrees of freedom $\bar{\bm{u}}_B(t)$. The enrichment space consists of an enrichment function $\psi({\bm{x},t)}$ times a polynomial of degree $l$:
\begin{equation}
\mathcal{V}_h^{\tilde{\bm{u}}} = \{\tilde{\bm{u}}_h \in (L^2)^d : \tilde{\bm{u}}_h |_{\Omega_e} \in (\psi P_l(\Omega_e)), \forall e \in \tilde{\Omega}_h \}.
\end{equation}
Written in terms of shape functions, the enrichment expansion reads
\begin{equation}
\tilde{\bm{u}}_h(\bm{x},t)=\psi(\bm{x},t)\sum_{B \in N^{l}} N_{B}^{l}(\bm{x})\tilde{\bm{u}}_{B}(t).
\label{eq:enrichment}
\end{equation}
Usually it is sufficient to enrich a single layer of cells in the vicinity of the no-slip boundary, $\tilde{\Omega}_h\subset \Omega_h$, and we set $\tilde{\bm{u}}_h(\bm{x},t)=\bm{0}$ outside this area. The enrichment function represents an a priori known approximate solution and the definition applied in this article is discussed in the following subsection. The shape functions $N^{l}$ employed here may be chosen independently to the ones used in~\eqref{eq:std_fe} and we take linear shape functions with $l=1$ throughout this article in order to keep the additional degrees of freedom introduced in~\eqref{eq:enrichment} to a minimum. The weighting of the wall function with a linear FE space gives the method high flexibility in adapting the enrichment to separated flow conditions.

The composition of the resulting function space is illustrated in Figure~\ref{fig:decomp} for an equilibrium boundary layer and for a non-standard boundary layer profile, such as in separated flows. In an equilibrium boundary layer, the velocity profile may to the largest extent be captured by the enrichment while the full flexibility of a high-order polynomial plus a linearly weighted wall function are available if the solution demands.

In the comparison of the enrichment formulation employed in~\cite{Krank16} within the continuous Galerkin method, we have taken into account two simplifications. Blending of function spaces at the interface between enriched and non-enriched elements is automatically included in the discontinuous Galerkin method since the weak coupling between the elements allows non-conforming shape functions, making the ramp functions used in~\cite{Krank16} unnecessary. Further, we do not subtract the enrichment function by its nodal values to simplify implementation using common FE libraries. As a consequence, the shape functions do not have to be modified on a basic level but are simply re-combined subsequently to existing evaluation routines (see Section~\ref{sec:impl} for a possible layout of implementation routines).

Finally, we discuss the treatment of the remaining variables $p$ and $\tilde{\nu}$. In typical boundary layers, high-gradient solutions do not occur in the pressure, thus we employ only the standard polynomial function space of degree $k$ to discretize $p_h\in\mathcal{V}_h^{p}$. The Spalart--Allmaras model further results in a linear distribution of the eddy viscosity $\tilde{\nu}\sim y^+$ between the wall and the outer edge of the logarithmic layer, again under equilibrium assumptions~\cite{Kalitzin05}. We therefore assume that a polynomial element of degree $k$ adds sufficient flexibility to $\tilde{\nu}_h\in\mathcal{V}_h^{\tilde{\nu}}=\mathcal{V}_h^{p}$ in order to capture more general profiles of $\tilde{\nu}$ in non-equilibrium boundary layers as well.

\subsection{Enrichment function}
\label{sec:spalding}
As an enrichment function, it would be possible to employ any wall function including both the viscous sublayer and the logarithmic region in one $C^1$-continuous formula that fulfills the no-slip boundary condition $\bm{u}(y=0)=0$ and $\frac{\partial u^+}{\partial y^+}|_{y=0}=1$. The latter requirement is essential for accurate predictions of the wall shear stress~\cite{Dean76}. In particular, it would be possible to utilize the wall function implicitly given through the Spalart--Allmaras model using a procedure described in~\cite{Kalitzin05} or the advanced wall functions in~\cite{Knopp06,Popovac07}. Considering our very positive experiences with this function~\cite{Krank16}, we use a minor modification of Spalding's law~\cite{Spalding61}, given as
\begin{equation}
y^+=\frac{\psi}{\kappa}+e^{-\kappa B}\left(e^{\psi}-1-\psi-\frac{\psi^2}{2!}-\frac{\psi^3}{3!}-\frac{\psi^4}{4!}\right)
\label{eq:spald}
\end{equation}
with $y^+=\frac{y}{\nu}\sqrt{\frac{\tau_w}{\rho}}$, where the wall shear stress is denoted $\tau_w$ and the density $\rho$. The von K\'arm\'an constant $\kappa = 0.41$ and $B=5.17$ according to Dean~\cite{Dean78} are used.

The wall shear stress is chosen as a function in space and time in order to take into account the local features of the flow as well as their temporal evolution in the model. In order to facilitate the evaluation of the enrichment, we define a discrete wall distance $y_h$ and a wall shear stress $\tau_{w,h}$ in terms of a finite element expansion as in~\cite{Krank16}. For this FE expansion we choose linear continuous shape functions of degree $m=1$. The wall shear stress may then be computed on each node in $N^{c,m}$ with
\begin{equation}
\tau_{w,B}=\frac{\int_{\partial \Omega^D} N_B^{c,m}(\bm{x})\rho \nu\frac{\partial \bm{u}_h}{\partial y}\big|_{y=0} dA}{\int_{\partial \Omega^D} N_B^{c,m}(\bm{x}) dA}
\end{equation}
and interpolated using
\begin{equation}
\tau_{w,h}=\sum_{B \in N^{c,m}}N_B^{c,m} \tau_{w,B}.
\end{equation}
This choice allows representation of a varying $\tau_w$ within the elements in contrast to for example a constant distribution (DG of degree $m=0$).
In order to enhance the robustness of the method for small values of $\tau_w$, for example at reattachment locations, a minimum of $2\%$ of the average value is specified similar to~\cite{Krank16}. The wall shear stress is further re-computed in every time step to allow for temporal adaptivity and the velocity is projected onto the new FE space via $L^2$ projection again according to~\cite{Krank16}. The first derivatives of the enrichment shape functions, which are part of the weak forms presented below, are computed with the formula included in~\cite{Krank16}.

\section{Application to a high-order DG code}
\label{sec:num}
The present wall modeling approach may be embedded in any steady RANS or unsteady URANS solver based on the discontinuous Galerkin method. In this paper we exemplify the enrichment with our high-performance matrix-free Navier--Stokes solver INDEXA employing high-order discontinuous Galerkin discretizations~\cite{Krank16b}. This code is complemented by the Spalart--Allmaras model in the current work.

In the first subsection (Section~\ref{sec:mat}), the scheme is summarized by a matrix formulation, paving the way for a discussion of the implementation of the enrichment approach. The full temporal discretization and spatial Galerkin formulation are included in Appendices~\ref{sec:temp} and~\ref{sec:gal} for the interested reader.

\subsection{Discrete formulation of a temporal splitting scheme}
\label{sec:mat}
The Navier--Stokes equations are integrated in time using the high-order semi-explicit velocity-correction scheme by Karniadakis et al.~\cite{Karniadakis91}, which consists of three sub-steps. The convective term is first treated explicitly. A Poisson problem is subsequently solved for the pressure and the latter is applied to obtain a divergence-free velocity in a local projection. Finally, the viscous term is accounted for in a Helmholtz-like equation. This time stepping algorithm is augmented by an explicit Spalart--Allmaras step in the present article avoiding nonlinear iterations entirely. The temporal discretization resulting from the strong forms~\eqref{eq:conti}--\eqref{eq:mom} is described in Appendix~\ref{sec:temp}.

Weak forms are derived as usual by multiplication of the temporally discretized equations with appropriate weighting functions, integration over one element volume, and integration by parts to obtain flux forms. Inserting the definitions of $\bm{u}_h$, $p_h$, and $\tilde{\nu}_h$ including the enrichment from Section~\ref{sec:wm} yields the final Galerkin formulations. These are very similar to the ones presented in~\cite{Krank16b} for the Navier--Stokes equations and are fully detailed in Appendix~\ref{sec:gal} for all sub-steps of the scheme. Herein, the Lax--Friedrichs numerical flux is used to discretize the convective terms in the Spalart--Allmaras equation and the momentum equation, as well as penalty methods for the second derivatives in the diffusive term of the Spalart--Allmaras equation, the Poisson equation, and the viscous term. To this end, the symmetric interior penalty method~\cite{Arnold82} is used for the non-enriched variables while the non-symmetric version~\cite{Riviere99} is preferred for the viscous step where the enriched velocity is the solution variable. The motivation for this choice is that the non-symmetric interior penalty method is known to be stable with much lower requirements on the penalty parameter~\cite{Riviere99}. This is beneficial in our solver since a general interior penalty parameter definition for non-polynomial shape functions is not straight forward. For the diffusive and viscous terms, we employ harmonic weighting of the discontinuous material property $\tilde{\nu}_h$~\cite{Burman12} in order to enable a stable scheme even for significant discontinuities in the eddy viscosity variable.

All integrals in the Galerkin formulation are evaluated with appropriate quadrature rules according to~\cite{Krank16b} for the polynomial elements. The numerical integration of enriched elements requires special attention due to the non-polynomial shape functions. As discussed in~\cite{Fries10}, it may be efficient to construct problem-tailored quadrature formulas for such applications, especially regarding the high-gradient direction, i.e. the wall-normal direction. For simplicity, we employ Gaussian quadrature for all elements and increase the number of points used within the enriched elements as in~\cite{Krank16}. Typically, at least 15 quadrature points are necessary in the wall-normal direction, and the requirement increases if the enriched element spans a wider range in $y^+$-units.

Upon evaluation of all integrals in the Galerkin formulation, one arrives at the matrix formulation, which is used in the following to summarize the numerical method and serves as a basis for discussion of the implementation in the second subsection.

The sub-steps are:

\paragraph{Explicit Spalart--Allmaras step}
In the first sub-step, the Spalart--Allmaras transport equation is explicitly advanced in time, avoiding costly non-linear iterations regarding both $\tilde{\nu}$ and $\bm{u}$. The temporal derivative is treated by a backward-differentiation formula (BDF) of order $J$ and the right-hand side is extrapolated from the previous time steps with an extrapolation formula (EX) of the same accuracy. We get
\begin{multline}\label{eq:samat}
\gamma_0 \tilde{\bm{N}}^{n+1} = \sum_{i=0}^{J-1}\alpha_i \tilde{\bm{N}}^{n-i}\\
- \Delta t  \tilde{\bm{M}}^{-1} \sum_{i=0}^{J-1}\beta_i \left(\tilde{\bm{F}}^c\left(\bm{U}^{n-i}\right)\tilde{\bm{N}}^{n-i}-\tilde{\bm{F}}^{\tilde{\nu}}\left(\tilde{\bm{N}}^{n-i}\right)\tilde{\bm{N}}^{n-i}-\bm{Q}\left(\bm{U}^{n-i},\tilde{\bm{N}}^{n-i}\right)\right)
\end{multline}
with the eddy viscosity vector $\tilde{\bm{N}}$, velocity vector $\bm{U}$, mass matrix $\tilde{\bm{M}}$, convective as well as diffusive flux terms $\tilde{\bm{F}}^c$ and $\tilde{\bm{F}}^{\tilde{\nu}}$, and the source terms $\bm{Q}$. The increment in time is denoted $\Delta t$, $n$ is the time step number, and $\alpha_i$, $\beta_i$, as well as $\gamma_0$ are time integrator coefficients according to~\cite{Karniadakis91} for constant time step sizes. As we consider temporally varying time step increments, the adaptation of these coefficients is necessary in every time step.
The explicit formulations of both the convective and the diffusive term in this equation restrict the time step size by the Courant--Friedrichs--Lewy (CFL) condition and the diffusion number $D$. An adaptive algorithm maximizing the time step size while fulfilling both conditions is described in Appendix~\ref{sec:dt_algo}.
\paragraph{Explicit convective step}
The Navier--Stokes equations are integrated in time as usual with the present velocity-correction method. The convective term is first integrated explicitly with the same scheme as the Spalart--Allmaras equation yielding the vector of the first intermediate velocity $\hat{\bm{U}}$:
\begin{equation}\label{eq:convmat}
\gamma_0 \hat{\bm{U}} = \sum_{i=0}^{J-1}\alpha_i \bm{U}^{n-i}- \Delta t  \bm{M}^{-1} \sum_{i=0}^{J-1}\beta_i \bm{F}^c\left(\bm{U}^{n-i}\right)\bm{U}^{n-i}+\bm{F}^{n+1},
\end{equation}
where $\bm{M}$ is the mass matrix, $\bm{F}^c$ is the convective flux, and $\bm{F}$ the body-force vector. This step is also subject to the CFL condition due to its explicit character, similarly to the Spalart--Allmaras step.
\paragraph{Pressure Poisson equation and projection}
The pressure is subsequently computed by solving a Poisson equation, which is the first global problem to be solved and given as
\begin{equation}\label{eq:poissonmat}
\bm{L} \bm{P}^{n+1} = \frac{\gamma_0}{\Delta t}\bm{A}\hat{\bm{U}},
\end{equation}
with $\bm{L}$ the Laplace operator, $\bm{P}$ the pressure vector, and $\bm{A}$ the divergence operator. 
Using the pressure available, a divergence-free velocity vector $\hat{\hat{\bm{U}}}$ may be obtained by a projection:
\begin{equation}\label{eq:project}
\left(\bm{M}+ \bm{D}\right) \hat{\hat{\bm{U}}} = \left( \bm{M}\hat{\bm{U}} + \frac{\Delta t}{\gamma_0} \bm{B} \bm{P}^{n+1}\right)
\end{equation}
with the gradient operator $\bm{B}$. In our formulation, the projection step is augmented by a div-div penalty term summarized in matrix $\bm{D}$, which is used to stabilize the scheme by enhancing mass conservation properties~\cite{Krank16b}, see Appendix~\ref{sec:gal} for details. This projection is a local problem since both the mass matrix and the div-div matrix are block-diagonal.
\paragraph{Implicit viscous step}
Finally, we arrive at the velocity solution at time $t^{n+1}=t^n+\Delta t$ by solving the second global equation system, consisting of the Helmholtz equation, given as
\begin{equation}\label{eq:viscous}
\left(\frac{\gamma_0}{\Delta t}\bm{M}-\bm{F}^{\nu}\left(\tilde{\bm{N}}^{n+1}\right)\right)\bm{U}^{n+1} = \frac{\gamma_0}{\Delta t}\bm{M} \hat{\hat{\bm{U}}},
\end{equation}
where the viscous flux is denoted $\bm{F}^{\nu}$.

\subsection{Implementation}
\label{sec:impl}
The matrix formulation in the previous subsection consists of cell and face integrals emerging from the weak forms. For these evaluations, we employ the high-performance computational kernels by Kronbichler and Kormann~\cite{Kronbichler12} within the open-source \texttt{deal.II} finite element library~\cite{Bangerth16} implemented in the C++ programming language. In this framework, the evaluation of the finite element interpolation in a quadrature point $\bm{u}_h(\bm{x}_q,t)$ as well as multiplication by test functions $\bm{v}_h(\bm{x}_q,t)$ and summation over quadrature points is provided by a class called \texttt{FEEvaluation}. The computational kernels include read (gather) and write (scatter) operations into global vectors, evaluation and integration routines based on sum factorization, as well as the combination of values and gradients on quadrature points.

For the evaluation of enriched function spaces, a modular extension to \texttt{FEEvaluation} has been developed. As this extension can be used generically and is not restricted to the current setting, it is detailed in the following paragraph. According to the definitions in \eqref{eq:space} and \eqref{eq:enrichment}, the evaluation of the enriched function $\bm{u}_h(\bm{x}_q,t)$ combines the interpolation of a standard polynomial space $\bar{\bm u}_h$ of degree $k$ with the interpolation $\tilde{\bm{u}}_h$ of degree $l$. For the combined evaluation according to Equation \eqref{eq:space}, the two polynomial representations underlying $\bar{\bm{u}}_h$ and $\tilde{\bm{u}}_h$ are each evaluated in the quadrature point location $\bm{x}_q$ with index $q$. The following C++ code shows the implementation of the function that computes the enriched interpolation in the $q$-th quadrature point, \texttt{get\_value}, where the two components are combined and the enrichment function is multiplied. The second method described here concerns integration where the action on a quadrature point $q$ is to \emph{submit} a value prior to the multiplication by all the test functions in the quadrature point, \texttt{submit\_value} \cite{Kronbichler12,Bangerth16}. Note that the value to be tested is submitted to both the test function slot associated with the polynomial function space \texttt{function\_space\_1} and the polynomial slot of the enrichment polynomials \texttt{function\_space\_2} so that each receive a contribution in the respective degrees of freedom in the residual. After the loop over quadrature points, the actual multiplication by all basis functions and summation over all basis functions via sum factorization is done in a function called \texttt{integrate}.

\begin{lstlisting}[language=C++]
template <...> class EnrichedEvaluation
{
  typedef typename FEEvaluation<...>::value_type  value_type;
  typedef typename FEEvaluation<...>::scalar_type scalar_type;

  value_type get_value(const unsigned int q) const
  {
    return function_space_1.get_value(q) +
           enrichment_function[q] * function_space_2.get_value(q);
  }

  void submit_value (const value_type value_to_test,
                     const unsigned int q)
  {
    function_space_1.submit_value(value_to_test, q);
    function_space_2.submit_value(enrichment_function[q]*value_to_test, q);
  }
  ...
  FEEvaluation<...> function_space_1;
  FEEvaluation<...> function_space_2;
  scalar_type *enrichment_function;
};
\end{lstlisting}

The implementation of \texttt{FEEvaluation} uses templates on the space dimension, polynomial degree, the number of integration points, and number of components that are omitted for brevity. For the fluid velocity, the type \texttt{value\_type} denotes a tensor with $d$ components but the same code can be used for scalar enrichments when the \texttt{value\_type} is a scalar. Furthermore, the particular implementation in \texttt{deal.II} combines the evaluation of several elements at once for making use of SIMD instructions (vectorization) in modern CPUs \cite{Kronbichler12}, which is why the inner quantity in a component of the tensor is not simply a \texttt{double} field but rather a short array of \texttt{double} variables.

In the evaluator, Spalding's law is a scalar quantity of type \texttt{scalar\_type} that is accessed via a pointer \texttt{enrichment\_function} to a table of the values on all enriched cells and all quadrature points. This factor is pre-computed prior to each time step using another \texttt{FEEvaluation} evaluator accessing the continuous finite element vectors of degree $m$ for $\tau_{w,h}$ and $y_h$ and resolving the formula for $\psi(\bm{x}_q)$ \eqref{eq:spald}. The latter quantity, Spalding's implicitly given wall function, is evaluated numerically with the algorithm described in~\cite{Krank16}. In a similar manner, the gradients in quadrature points and the testing by gradients of the test functions are a combination of the gradients on the individual fields, including the chain rule in the enriched part that also involves the gradient of Spalding's law with respect to the $\bm{x}$ coordinates according to the derivation in~\cite{Krank16}.

In the case of continuous FE spaces underlying the enrichment, the blending of function spaces could be realized via constraint matrices in \texttt{deal.II}, as detailed in~\cite{Davydov16}. As aforementioned, blending is automatically handled by the discontinuous Galerkin method through the weak formulation of the element coupling.

This enrichment evaluator \texttt{EnrichedEvaluation} is included in our existing Navier--Stokes code INDEXA~\cite{Krank16b} that also contains standard polynomial code paths where the additional operations due to the enrichment are undesirable. In order to avoid re-implementing all the weak forms of the Navier--Stokes equations with different evaluators, the generic programming capabilities of the C++ programming language are used via templates. To this end, a wrapper class \texttt{FEEvaluationWrapper} is introduced that contains a template argument to switch between a basic \texttt{FEEvaluation} object in standard simulations and \texttt{EnrichedEvaluation} in simulations including enriched elements. Both classes have the same interface, allowing for a seamless integration into the solver with simply a change in types and zero computational overhead in non-enriched simulations.

The FE operators enable matrix-free implementations of all sub-steps of the present scheme, including the pressure Poisson equation and the viscous Helmholtz problem; the detailed solution procedures for all steps are given in~\cite{Krank16b}. In particular, we use inverse mass preconditioners for the local projection solver as well as the viscous solver. Herein, the inverse mass matrix $\bm{M}^{-1}$ on enriched elements is the only variable in our algorithm which cannot be evaluated in a matrix-free manner. Since the mass matrix is block-diagonal, the inverse may be calculated on each element independently. Due to repeated application, we pre-compute a scalar mass matrix on each element ahead of every time step and use an LU factorization for applying the action of its inverse on each velocity component. Finally, the standard solution procedures according to~\cite{Krank16b} for all iterative solvers give similar iteration counts compared to the standard polynomial case. From this fact we conclude that conditioning is not an issue.

\begin{table}
\caption{Channel flow cases and resolutions. $N_{e,i}$ number of elements per spatial direction $i$, $Re_{\tau}$ friction Reynolds number, $\gamma$ mesh stretching parameter, and $\Delta  y_{1e}^+$ width of first off-wall element. All computations employ $k=4$ polynomial degree of standard space, $l=1$ polynomial degree of enrichment space as well as a single enriched element row at the wall.}
\label{tab:ch_flows}
\begin{tabular*}{\textwidth}{l @{\extracolsep{\fill}} l l l l l l l}
\hline
Case     & $N_{e,1} \times N_{e,2}$ &$Re_{\tau}$   & $\gamma$ & $\Delta  y_{1e}^+$
\\ \hline \noalign{\smallskip}
$ch\_N8^2\_k4l1$   & $8 \times 8$    & $180$& - & $45$ \\
                   & $8 \times 8$    & $395$& - & $99$\\
                   & $8 \times 8$    & $590$& - & $118$\\
                   & $8 \times 8$    & $950$& - & $238$\\
                   & $8 \times 8$    & $2{,}000$& - & $500$\\
                   & $8 \times 8$    & $5{,}200$& - & $1{,}300$\\
                   & $8 \times 8$    & $10{,}000$& - & $2{,}500$\\
                   & $8 \times 8$    & $20{,}000$& - & $5{,}000$\\ \noalign{\smallskip}
$ch395\_N8^2\_k4l1$   & $8 \times 8$    & $395$& - & $99$\\
$ch395\_N16^2\_k4l1$   & $16 \times 16$    & $395$& - & $49$\\
$ch395\_N32^2\_k4l1$   & $32 \times 32$    & $395$& - & $25$\\ \noalign{\smallskip}
$ch50000\_N16^2\_k4l1$   & $16 \times 16$    & $50{,}000$& $2.0$ & $1{,}175$\\
$ch100000\_N16^2\_k4l1$   & $16 \times 16$    & $100{,}000$& $2.25$ & $1{,}664$\\
\hline
\end{tabular*}
\end{table}

\section{Numerical examples}
\label{sec:examples}
We investigate the present wall modeling approach using turbulent channel flow as well as flow past periodic hills. These two benchmark examples provide insight into the performance regarding wall-attached flow in the first case and separated flow with a high adverse pressure gradient in the second setup. All computations are carried out with a scheme of temporal accuracy of second order (BDF2) and we take the spatial polynomial degrees of $k=4$ for the standard velocity component, pressure, and eddy viscosity, as well as $l=1$ for the enrichment velocity component. This combination has performed best in terms of accuracy and execution speed in our preliminary investigations.

\begin{figure}[htb]
\centering
\includegraphics[trim= 0mm 0mm 0mm 0mm,clip,scale=0.22]{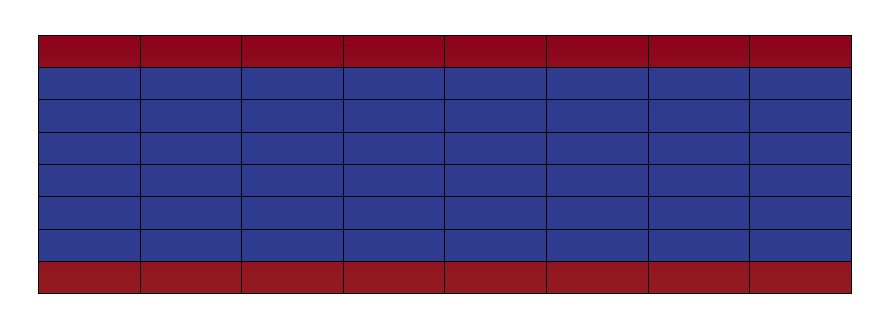}
\caption{Mesh for turbulent channel flow computations with $N=8^2$ elements. Enriched elements are colored red and standard polynomial elements blue, i.e. a single element layer is enriched at the walls.}
\label{fig:ch_mesh}
\center
\includegraphics[trim= 9mm 6mm 9mm 12mm,clip,scale=0.7]{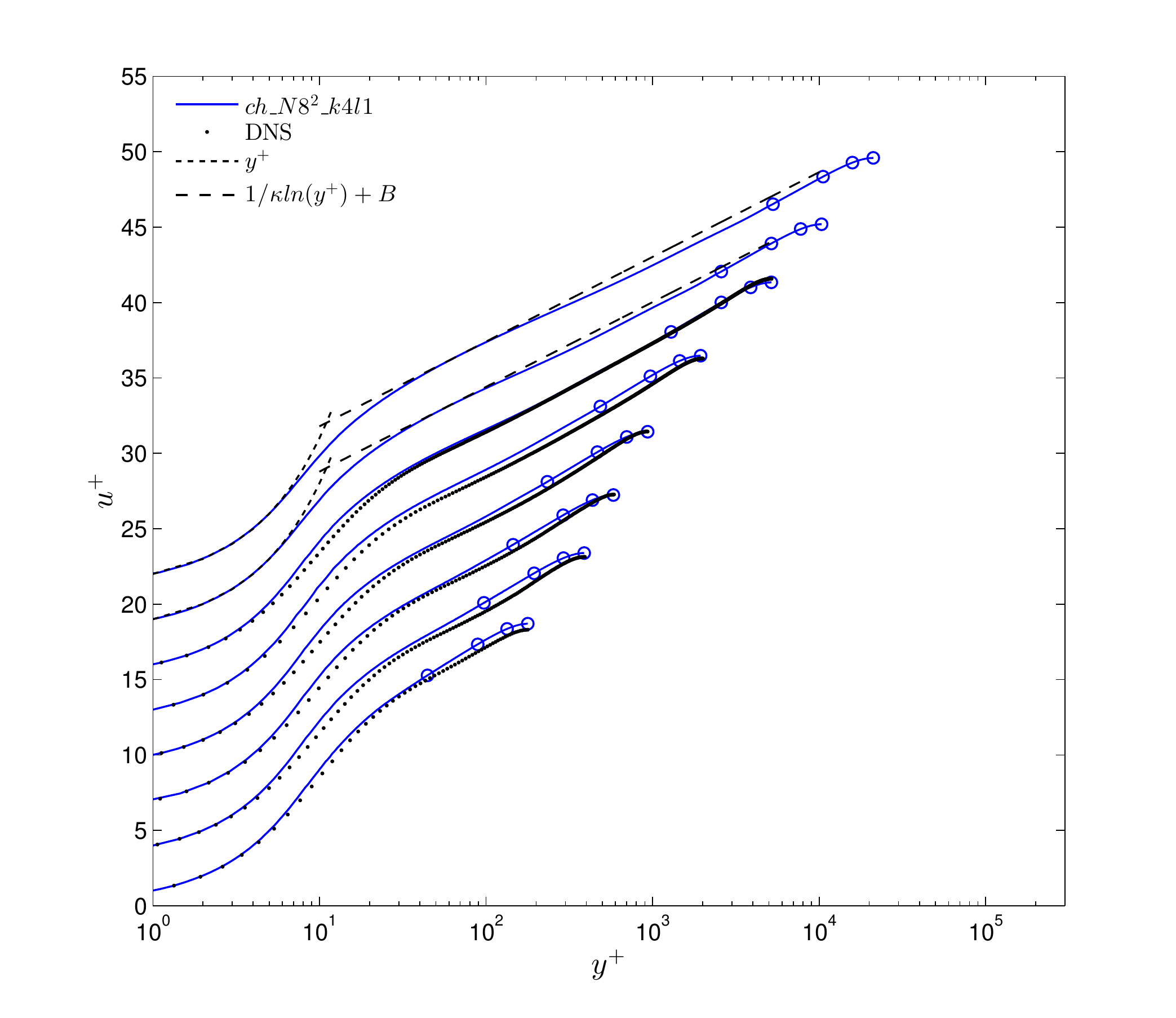}
\begin{picture}(100,0)
\put(156,306.5){\footnotesize $Re_{\tau}=20{,}000$}
\put(137.5,283){\footnotesize $Re_{\tau}=10{,}000$}
\put(119.5,262){\footnotesize $Re_{\tau}=5{,}200$}
\put(94,235){\footnotesize $Re_{\tau}=2{,}000$}
\put(75,208){\footnotesize $Re_{\tau}=950$}
\put(62.5,185.5){\footnotesize $Re_{\tau}=590$}
\put(52,165){\footnotesize $Re_{\tau}=395$}
\put(31.5,139){\footnotesize $Re_{\tau}=180$}
\end{picture}
\caption{Mean velocity of turbulent channel flow $u^+=u/{u_{\tau}}$ for $Re_{\tau}=180$, $395$, $590$, $950$, $2{,}000$, $5{,}200$, $10{,}000$ and $20{,}000$, each shifted upwards by three units for clarity. All computations have been carried out with the same mesh displayed in Figure~\ref{fig:ch_mesh} consisting of $N=8^2$ elements of fourth polynomial degree plus first degree for the enrichment within the first off-wall element row. Symbols indicate the location of element interfaces.}
\label{fig:ch_um_reydep}
\end{figure}

\subsection{Turbulent channel flow}
We consider turbulent flow in a plane channel of the dimensions $2\pi \delta \times 2\delta$ in streamwise and vertical direction, respectively, with channel-half width $\delta$. Periodic boundary conditions are specified in streamwise direction and no-slip boundary conditions are prescribed at the solid walls. All simulation cases and resolutions are presented in Table~\ref{tab:ch_flows}. The normalized mean velocity $u^+=u/u_{\tau}$ with $u_{\tau}=\sqrt{\tau_w/\rho}$ is post-processed at sufficiently many $y^+$ levels such that the full velocity profile may be compared to reference data. The following numerical experiments are separated into three groups investigating independence of the Reynolds number with the same mesh, mesh refinement with a constant Reynolds number, and application to high Reynolds numbers.

The first investigations discussed here employ all the same mesh as visualized in Figure~\ref{fig:ch_mesh} of $8 \times 8$ equally distributed cells and including a single enriched element row at the walls. With this spatial discretization, the enrichment constitutes only $4\%$ of the overall number of degrees of freedom. We perform simulations using friction Reynolds numbers $Re_{\tau}=u_{\tau}\delta/{\nu}$ in accordance with reference DNS by Moser et al.~\cite{Moser99} ($Re_{\tau}=180$, $395$, and $590$), Del {\'A}lamo and Jim{\'e}nez~\cite{Alamo03} ($Re_{\tau}=950$), Hoyas and Jim{\'e}nez~\cite{Hoyas06} ($Re_{\tau}=2{,}000$), as well as Lee and Moser~\cite{Lee15} ($Re_{\tau}=5{,}200$). Further friction Reynolds numbers of $Re_{\tau}=10{,}000$ and $Re_{\tau}=20{,}000$ are included and compared to the linear and log-laws in the respective $y^+$ regions. These simulation setups result in locations of the first off-wall element interface in a $y^+$ range between $y_{1e}^+=45$ and $5{,}000$ wall units, see Table~\ref{tab:ch_flows}. The results of the normalized mean velocity $u^+$ are depicted in Figure~\ref{fig:ch_um_reydep} and exhibit excellent agreement with reference data all the way down to the wall despite the significant difference in resolution respective wall units. In the middle of the channel, the mean velocity is over-predicted to a minor extent for $Re_{\tau}=180$ and marginally under-predicted for $Re_{\tau}=20{,}000$, the error is acceptable, however. A reason for this behavior may be that Spalding's law results in a slightly different velocity distribution in the buffer layer as compared to the Spalart--Allmaras model. Aiming at fine-tuning this approach, it would in future research certainly be valuable to investigate alternative enrichment functions, in particular the wall function implicitly given through the Spalart--Allmaras equation via a look-up table as suggested by~\cite{Kalitzin05}.

\begin{figure}[tb]
\center 
\includegraphics[trim= 9mm 0mm 9mm 7mm,clip,scale=0.7]{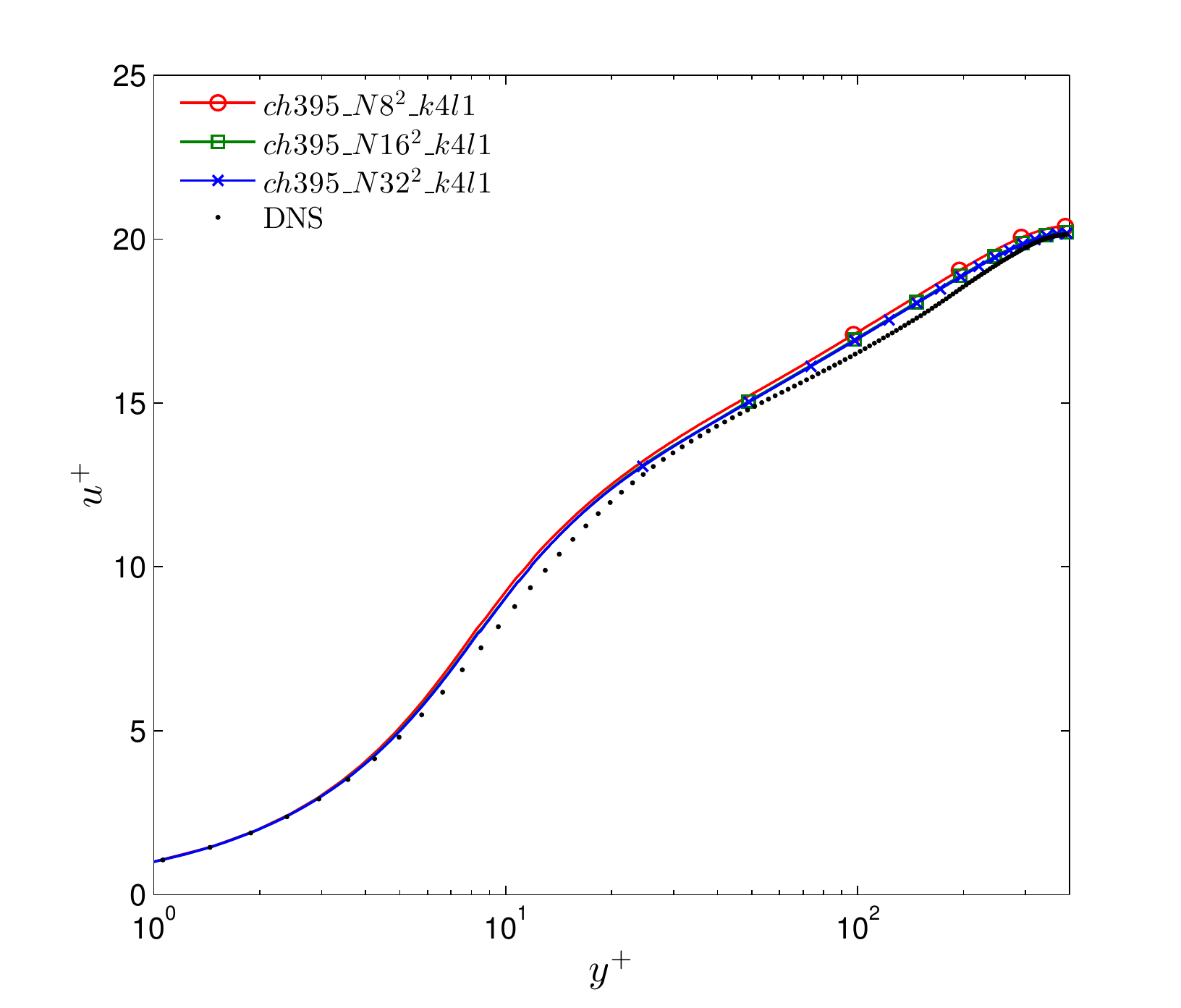}
\caption{Mean velocity of turbulent channel flow $u^+=u/{u_{\tau}}$ for $Re_{\tau}=395$ employing three different meshes of $N=8^2$, $N=16^2$ as well as $N=32^2$ elements. One single element layer closest to the wall is enriched in all cases. Symbols indicate the location of element interfaces.}
\label{fig:ch_um_mesh}
\end{figure}

Simple wall function approaches are often prone to inaccuracies during mesh refinement, especially when the coupling location moves inside the buffer layer. A refinement study is therefore performed for the case $Re_{\tau}=395$ using three successive refinement levels from $8 \times 8$ to $32 \times 32$ cells. The wall model is active only in the first off-wall cells, ranging up to $y_{1e}^+=99$ in the coarsest case and $y_{1e}^+=25$ in the finest case. The results displayed in Figure~\ref{fig:ch_um_mesh} show no difference between the three simulation cases as desired, although the wall model is switched off inside the buffer layer.

\begin{figure}[htb]
\center 
\includegraphics[trim= 9mm 0mm 9mm 7mm,clip,scale=0.7]{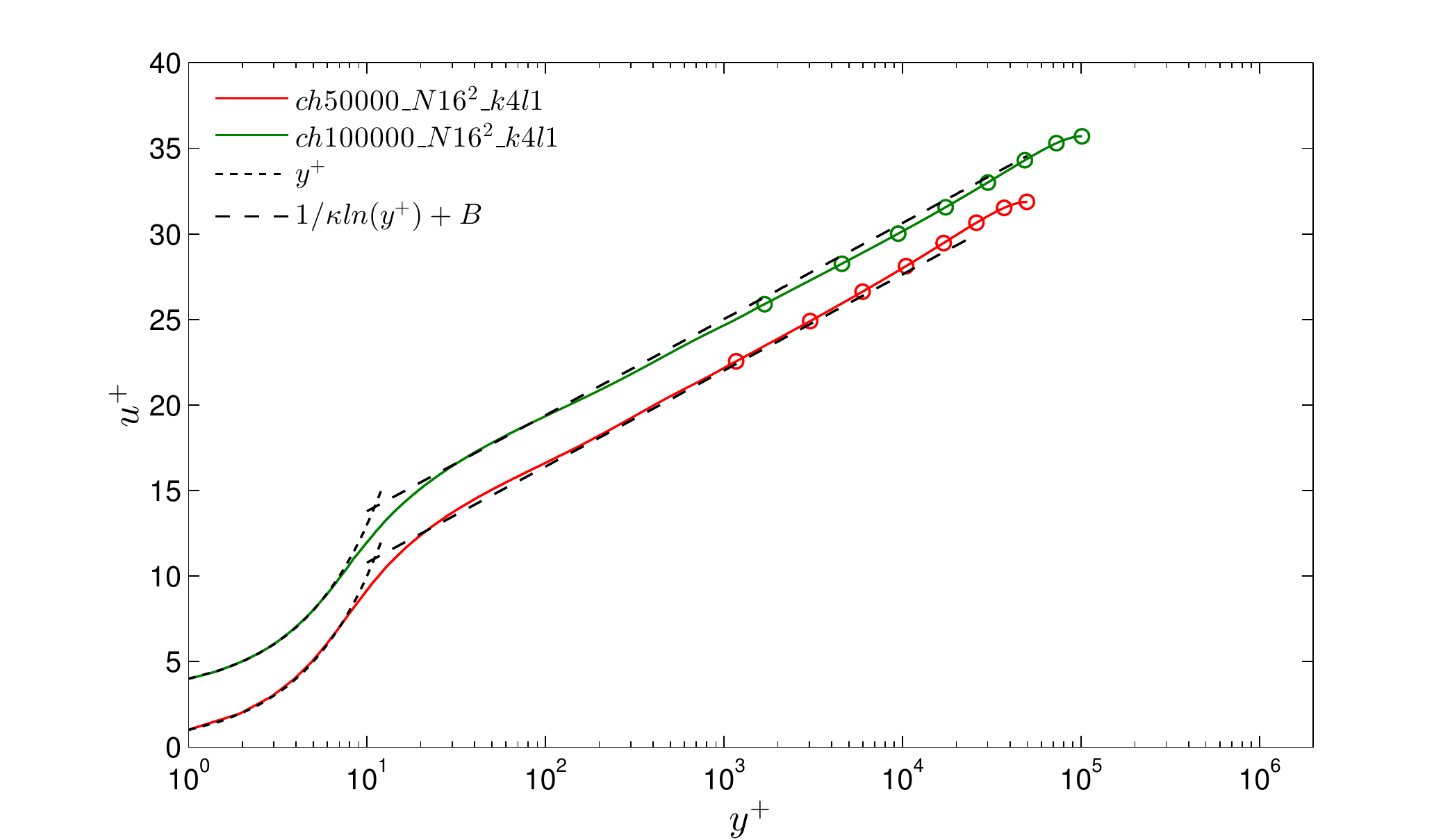}
\begin{picture}(100,0)
\put(158,215){\footnotesize $Re_{\tau}=100{,}000$}
\put(142,195.5){\footnotesize $Re_{\tau}=50{,}000$}
\end{picture}
\caption{Mean velocity of turbulent channel flow $u^+=u/{u_{\tau}}$ applied to high Reynolds numbers of $Re_{\tau}=50{,}000$ and $Re_{\tau}=100{,}000$, the latter shifted upwards by three units for clarity. One single element layer closest to the wall is enriched in both cases. Symbols indicate the location of element interfaces.}
\label{fig:ch_um_highre}
\end{figure}
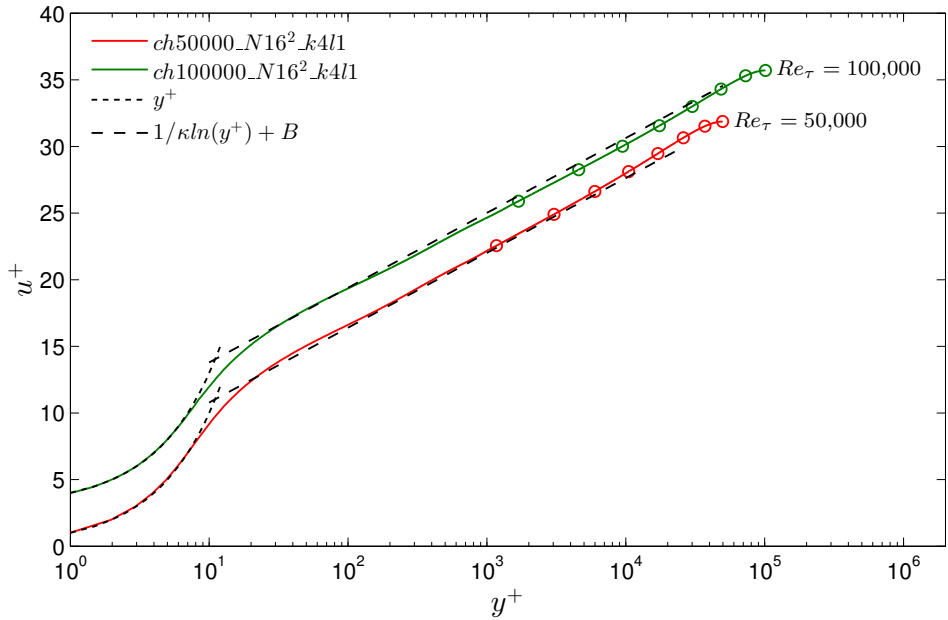

Many industrial applications, for example in the automotive, aerospace, or wind energy sector, demand for much higher friction Reynolds numbers. We demonstrate that higher Reynolds numbers may easily be computed with the present wall modeling approach. The friction Reynolds numbers $Re_{\tau}=50{,}000$ and $Re_{\tau}=100{,}000$ are investigated and compared to the linear and log-laws in the respective $y^+$ regions. Meshes using $16\times16$ cells are chosen and slightly refined towards the wall using a hyperbolic mapping of the form $f$: $[0,1] \to [-\delta, \delta]$:
\begin{equation}
x_2 \mapsto f(x_2)=\delta \frac{\tanh(\gamma (2x_2-1))}{\tanh(\gamma)}
\end{equation}
with the mesh stretching parameter $\gamma$ improving the resolution of the near-wall area. The values for $\gamma$ are selected as $2.0$ and $2.25$ according to Table~\ref{tab:ch_flows}. The results displayed in Figure~\ref{fig:ch_um_highre} confirm the accuracy of the present method and indicate suitability for high-Reynolds-number application areas.

From this section we draw the conclusion that wall modeling via function enrichment allows prediction of wall-attached flows with coarse meshes where the first cell spans $y^+$ bandwidths of up to $5{,}000$ wall units. The approach exhibits a high level of grid independence and the solution quality is retained also when increasing spatial resolution.

\begin{table}
\caption{Simulation cases and resolutions of the periodic hill benchmark. $Re_H=10{,}595$: $ph10595\_N16\times8\_k4l1$ coarse mesh with wall modeling, $ph10595\_N32\times16\_k4l1$ refined mesh with wall modeling, NTS\_Spalart\_Allmaras\_Ref reference using the Spalart--Allmaras model without wall modeling, FMRTL\_LES highly resolved LES. $Re_H=19{,}000$: $ph19000\_N16\times8\_k4l1$ coarse mesh with wall modeling, $ph19000\_N32\times16\_k4l1$ refined mesh with wall modeling, RM\_EXP experiments. Resolutions are specified in terms of elements per direction $N_{ei}$ and grid points $N_i \approx N_{ei}(k+1)$}
\label{tab:ph_flows} 
\begin{tabular*}{\textwidth}{l @{\extracolsep{\fill}} l l l l l}
\hline
Case     & $Re_H$ & Approach & $N_{e1} \times N_{e2} $ & $N_{1} \times N_{2}  \times N_{3} $ & $x_{1,reatt}/H$ \\ \hline
$ph10595\_N16\times8\_k4l1$  & $10{,}595$ & RANS (SA) & $16 \times 8$ & $80 \times 40$  & $7.32$ \\
$ph10595\_N32\times16\_k4l1$  & $10{,}595$  & RANS (SA) & $32 \times 16$ & $160 \times 80$ & $7.59$\\
NTS\_Spalart\_Allmaras\_Ref \hspace{0.05cm} & $10{,}595$ & RANS (SA) & - &$161\times161$  & $7.7$ \\
FMRTL\_LES  & $10{,}595$ & LES & - &$192 \times 128 \times 186$  & $4.6-4.7$ \smallskip\\ 
$ph19000\_N16\times8\_k4l1$ & $19{,}000$ & RANS (SA) & $16 \times 8$ & $80 \times 40$  & $7.35$ \\
$ph19000\_N32\times16\_k4l1$ & $19{,}000$ & RANS (SA) & $32 \times 16$& $160 \times 80$    & $7.59$\\
RM\_EXP & $19{,}000$ & experiment & - & - & $4.21$\\ 
 \hline
\end{tabular*}
\end{table}
\begin{figure}
\centering
\begin{minipage}[b]{0.497\linewidth}
\centering
\includegraphics[trim= 0mm 0mm 0mm 0mm,clip,width=1.\textwidth]{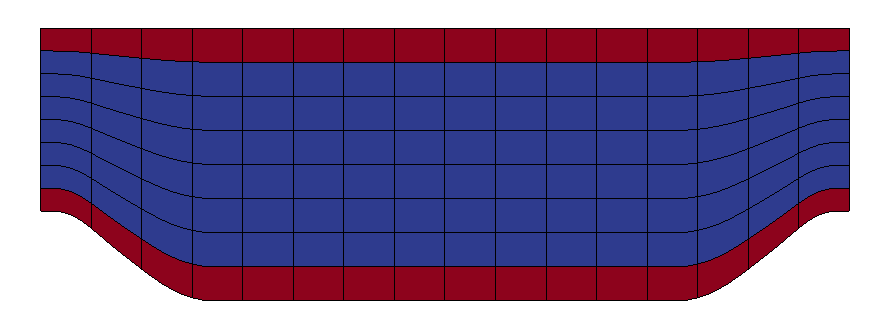}
\begin{minipage}{0.97\linewidth}
\centering
\caption{Grid of coarse refinement level. Enriched elements are colored red and
standard polynomial elements blue.}
\label{fig:ph_mesh}
\end{minipage}
\end{minipage}
\begin{minipage}[b]{0.497\linewidth}
\centering
\includegraphics[trim= 0mm 0mm 0mm 4mm,clip,scale=0.7]{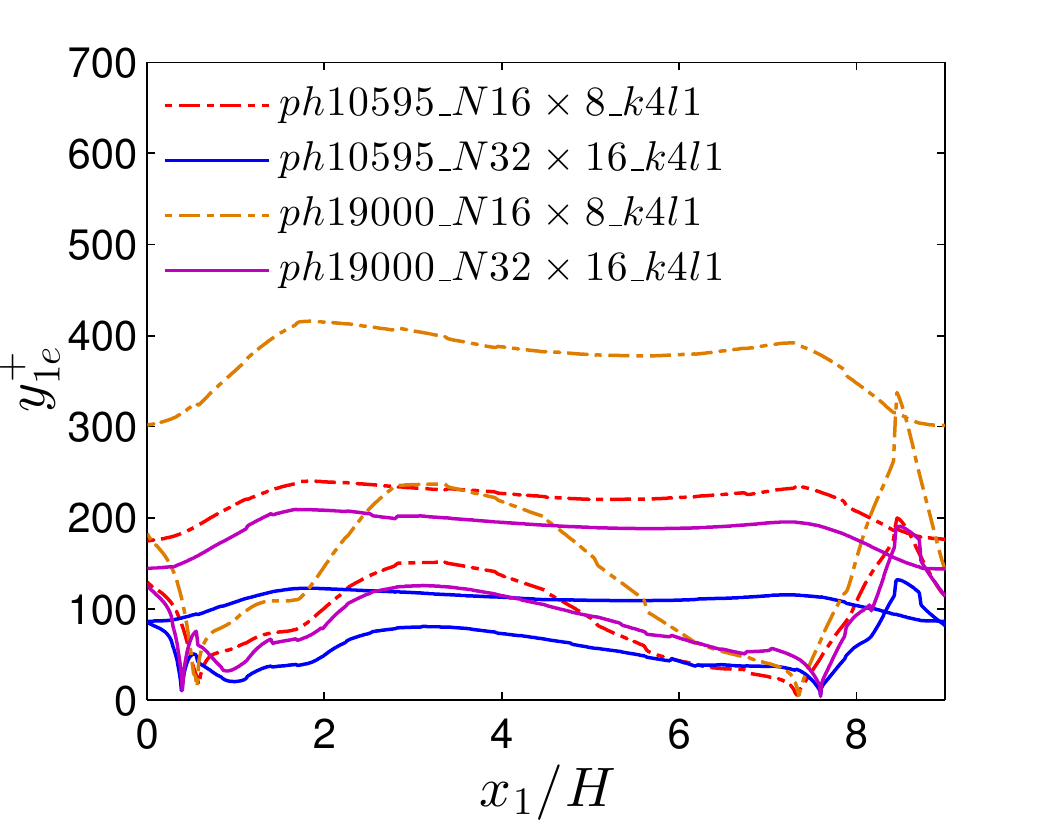}
\begin{minipage}{0.97\linewidth}
\centering
\caption{Location of first off-wall element interface $y_{1e}^+$ for flow past periodic hills. The shallower curves correspond to the upper wall.}
\label{fig:ph_y1}
\end{minipage}
\end{minipage}
\end{figure}
\subsection{Flow past periodic hills}
As a second numerical example we investigate flow past periodic hills according to Fr\"ohlich et al.~\cite{Frohlich05} at a Reynolds number based on the hill height $H$ of $Re_H=10{,}595$ and according to Rapp and Manhart~\cite{Rapp11} at $Re_H=19{,}000$. This example is challenging for common wall modeling approaches since it includes a separation bubble as well as a high adverse pressure gradient, violating the assumptions inherent to equilibrium wall modeling approaches. Wall modeling via function enrichment on the contrary has exhibited an outstanding performance in~\cite{Krank16} in the context of wall-modeled LES. This example has further been investigated by many researchers, regarding RANS in particular within the European initiative ``Advanced Turbulence Simulation for Aerodynamic Application Challenges'' (ATAAC) as well as Jakirli\'c and co-workers~(see, e.g.,~\cite{Jakirlic15}).

The domain of the present computations is of the dimensions $9H\times3.036H$ in streamwise and vertical direction, respectively, with no-slip boundary conditions at the top and bottom wall and periodic boundary conditions in horizontal direction.
Two meshes are considered, a coarser one using $16\times8$ cells and a finer one using $32\times16$ cells, both employing equidistant grid spacings for simplicity. The standard Galerkin component consists of polynomials of degree $k=4$ and the enrichment is based on polynomials of degree $l=1$ solely included within the first element layer near the no-slip walls. The additional degrees of freedom of the enrichment increase the overall degree-of-freedom count by no more than $4\%$ in the coarse case and $2\%$ in the fine case. All simulation cases including the labels used in the following plots are listed in Table~\ref{tab:ph_flows}. In order to enhance the representation of the curved boundary using such coarse meshes, the discretized geometry is mapped onto the reference curve by a polynomial of degree $k=4$ with facilities provided by the \texttt{deal.II} library~\cite{Bangerth16}. The resulting coarse mesh is displayed in Figure~\ref{fig:ph_mesh}. With these grids, the first off-wall elements including the enrichment span a range in $y^+$-units up to approximately $y_{1e}^+\approx 420$ at the top wall and $y_{1e}^+\approx 340$ at the hill crest for the coarser mesh as well as the higher Reynolds number. The detailed distribution of $y_{1e}^+$ for all cases is included in Figure~\ref{fig:ph_y1}.

Our results are compared to three data sets, which are also included in the overview in Table~\ref{tab:ph_flows}. Accurate reference data is provided by highly resolved LES of Fr\"ohlich et al.~\cite{Frohlich05} for the case $Re_H=10{,}595$ and by experiments of Rapp and Manhart~\cite{Rapp11} for $Re_H=19{,}000$. This reference data has been obtained from the ERCOFTAC QNET-CFD Wiki contributed by Rapp et al. \cite{RappERCOFTAC}. Further reference data considered is a wall-resolved simulation of the case $Re_H=10{,}595$ using the Spalart--Allmaras model, which has been conducted by Strelets and Adamian within the European ATAAC project and has been extracted from vector-graphics plots in~\cite{Jakirlic10}. Since we use the same RANS approach in this article, we expect convergence with refinement not to the more accurate reference data but to the latter RANS simulation.

\begin{figure}[htb]
\center 
\includegraphics[trim= 9mm 0mm 9mm 4mm,clip,scale=0.7]{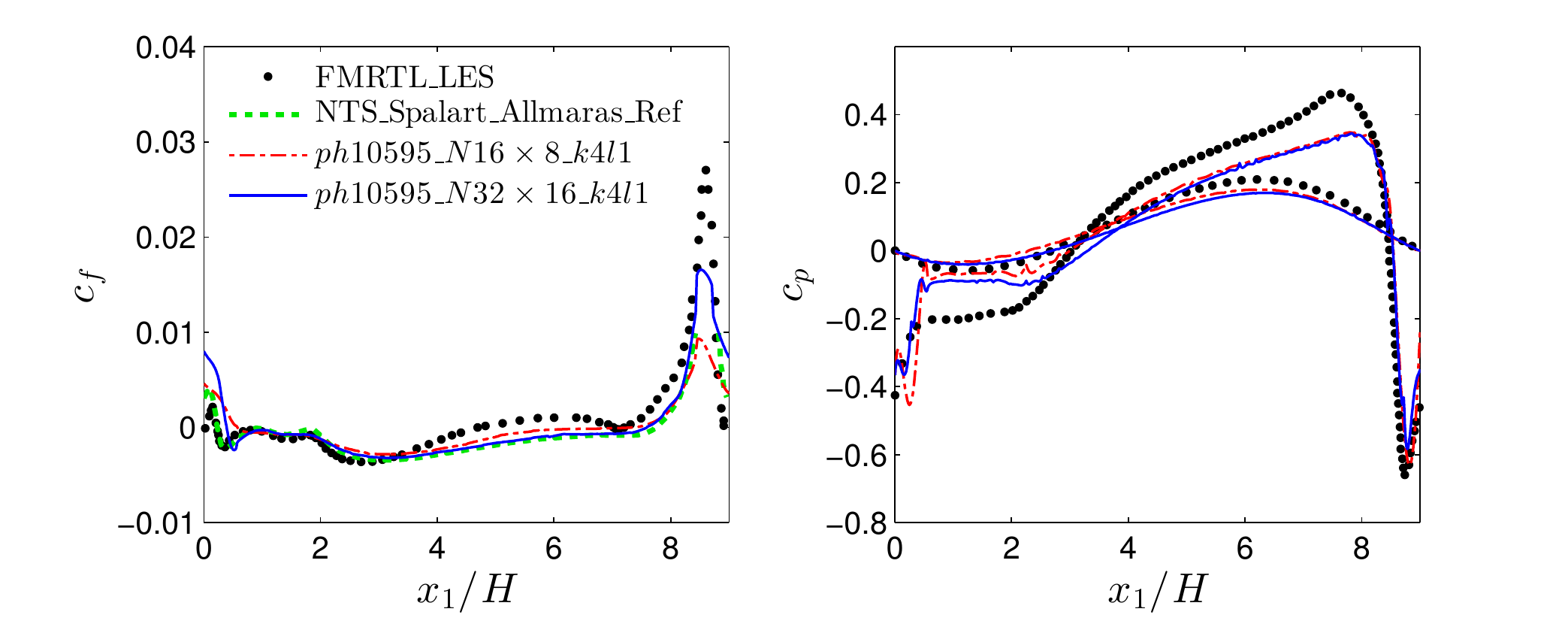}
\caption{Friction coefficient (left) and pressure coefficients (right) for $Re_H=10{,}595$. The shallower pressure curves corresponds to the top wall.}
\label{fig:ph_cfcp}
\end{figure}

We commence the discussion of the results with the friction and pressure coefficients $c_f$ and $c_p$ at the walls. They are defined as
\begin{equation*}
c_f=\frac{\tau_w}{\frac{1}{2} \rho u_b^2}\hspace{1cm}
c_p=\frac{p-p_{\mathrm{ref}}}{\frac{1}{2} \rho u_b^2}
\label{eq:cfcpdef}
\end{equation*}
with the bulk velocity $u_b$ and the reference pressure $p_{\mathrm{ref}}$ taken at $x_1=0$ on the upper wall. Results are plotted in Figure~\ref{fig:ph_cfcp} for $Re_H=10{,}595$. As for the skin friction, there is generally a very high level of agreement between the two meshes included in the present study, except at the hill crest, where the peak is much more pronounced in the fine case. It is remarkable that both curves are in very close accordance with the RANS reference data carried out with the Spalart--Allmaras model despite the coarseness of the meshes, again except in the vicinity of the hill, where significant deviations are visible. A comparison of the peak of the $c_f$ curve is unfortunately not possible with this reference data due to the lack of data points in this region. A further comparison with accurate reference data by well-resolved LES indicates that the general trend is predicted well, yet with a clear deviation from the reference. This discrepancy is also represented in the reattachment lengths $x_{1,reatt}/H$ included in Table~\ref{tab:ph_flows} which are significantly over-predicted by all computations using the present RANS model. The pressure curves according to Figure~\ref{fig:ph_cfcp} are of similar quality, since the difference between the two computations using the current approach is negligible while solely the general trend is predicted well.

\begin{figure}[t!]
\centering
\includegraphics[trim= 9mm 4mm 9mm 0mm,clip,scale=0.7]{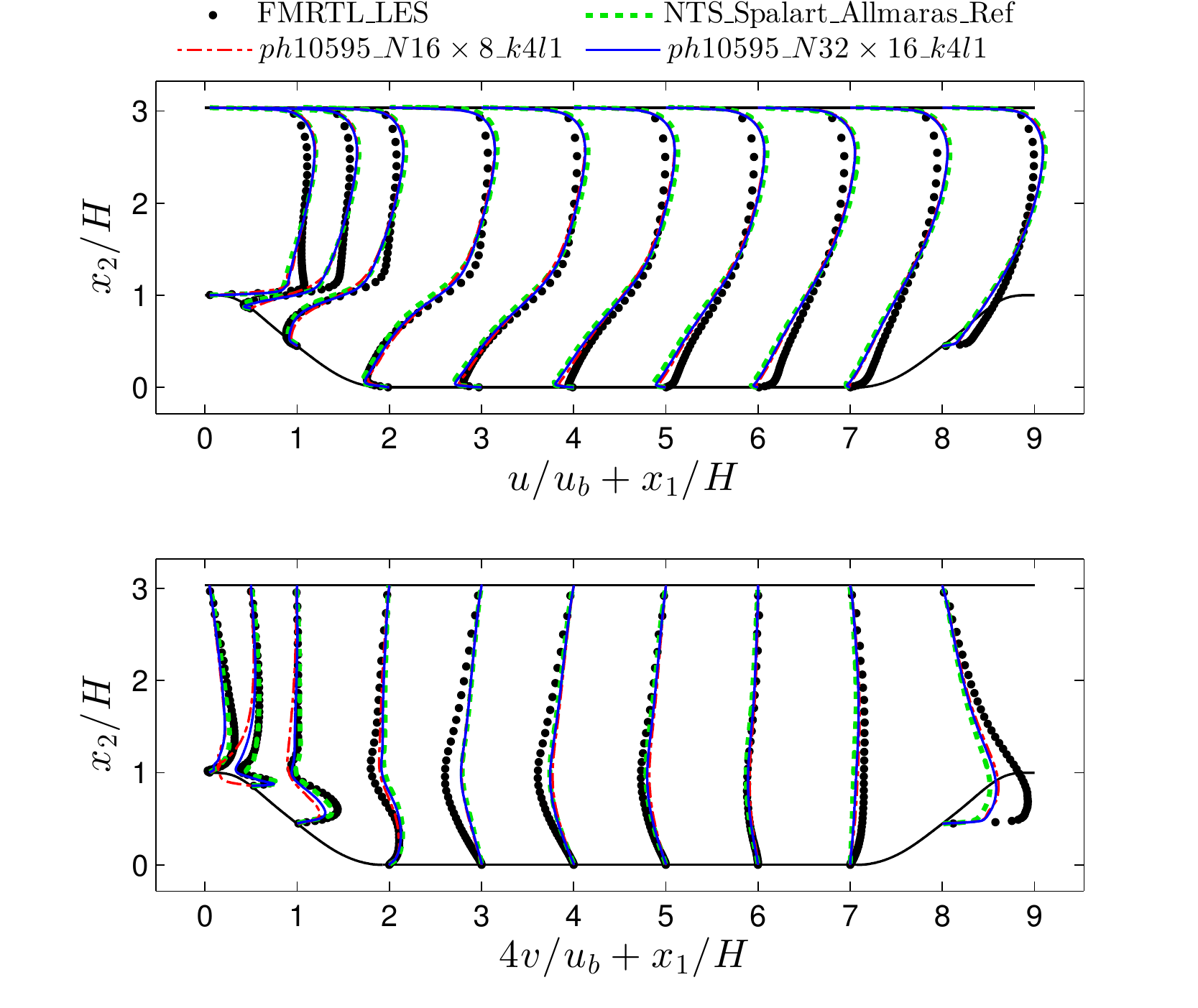}
\caption{Mean streamwise (top) and vertical (bottom) velocity components $u$ and $v$ of flow past periodic hills at $Re_H=10{,}595$.}
\label{fig:phum10595}
\vspace{5mm}
\includegraphics[trim= 9mm 4mm 9mm 5mm,clip,scale=0.7]{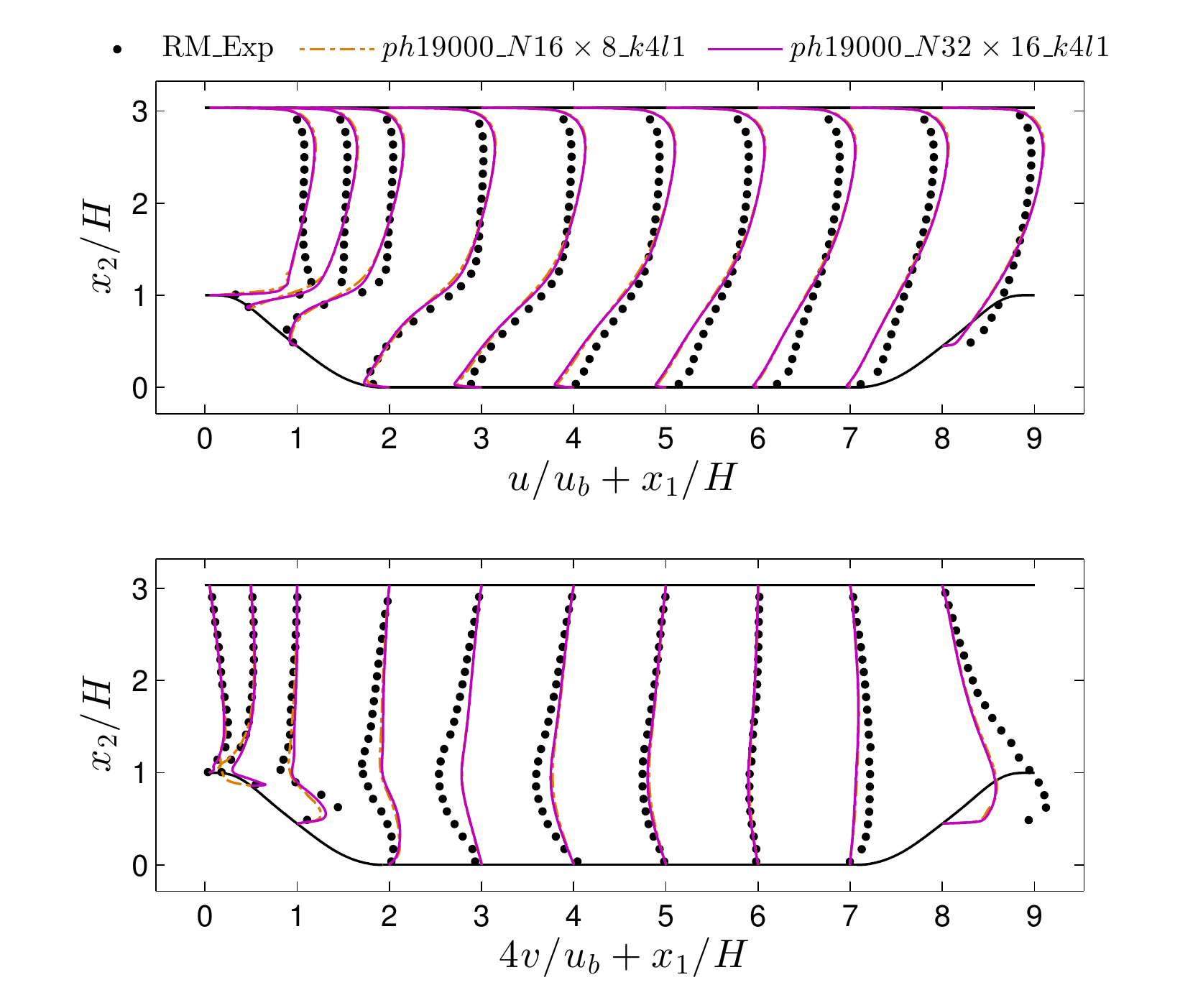}
\caption{Mean streamwise (top) and vertical (bottom) velocity components $u$ and $v$ of flow past periodic hills at $Re_H=19{,}000$.}
\label{fig:phum19000}
\end{figure}

Next, the mean velocities in streamwise and vertical direction $u$ and $v$ are plotted in Figure~\ref{fig:phum10595} at ten streamwise stations for the same Reynolds number $Re_H=10{,}595$. The two computations carried out with the present approach display only minor differences to the Spalart--Allmaras reference. This fact indicates that the wall model is well capable of providing an appropriate function space that yields results almost equal to considerably finer meshes without wall model. The mean streamwise velocity is in general predicted too high near the top wall whereas it is estimated too low near the bottom wall. As a consequence, the recirculation bubble is extended towards the hill on the right-hand side, which is in agreement with the observed over estimation of the reattachment length.

The application of the same meshes to the higher Reynolds number of $Re_H=19{,}000$ confirms the observations of the lower Reynolds number. Detailed reference data of the skin friction and pressure coefficients is not available for this case. The reattachment lengths included in Table~\ref{tab:ph_flows} are significantly over-predicted, however. The mean velocity displayed in Figure~\ref{fig:phum19000} shows very similar results to the lower Reynolds number case. Since refinement hardly leads to changes in the results, it may be concluded that the function space provided by the present wall model is appropriate and the remaining differences to reference data are due to the Spalart--Allmaras model.

In summary, it has been found that the present wall modeling approach provides an appropriate function space even in separated flows, as we have been able to reproduce reference results using the Spalart--Allmaras model without wall model. Moreover, it has been shown that the Spalart--Allmaras model yields deficient results for this flow example. A major benefit of our wall model is that refinement does not lead to a degradation of the result quality as opposed to common wall models, since the polynomial component of the elements is retained and automatically jumps in if necessary.

\section{Conclusion}
\label{sec:con}
In this paper we have presented a wall modeling approach for RANS which overcomes the common limitations of grid dependence and poor performance in separated flows. This is achieved by solving the full Navier--Stokes equations including the Spalart--Allmaras model with exact wall boundary conditions. The method is capable of employing grids where the first off-wall cell spans a $y^+$ range of up to 5,000 wall units. The key ingredient of this approach is that the Galerkin function space is enriched with a wall function in addition to the usual polynomial shape functions. The numerical method then tries to make optimal use of these contributions.

In flow past periodic hills, the results delivered by the Spalart--Allmaras model are not fully satisfactory. The straight-forward extension of the present scheme towards hybrid RANS/LES~\cite{Spalart97} would therefore be highly interesting. We believe that the LES capabilities of the code~\cite{Krank16b} in conjunction with the present wall modeling approach will form a powerful computational tool for future applications.

\appendix
\section{Spalart--Allmaras model}
\label{sec:sa_sourceterm}
The source terms of the Spalart--Allmaras model are considered in the following form assuming a fully turbulent flow:
\begin{equation}
Q(\bm{u},\tilde{\nu}) = c_{b1}\tilde{S}\tilde{\nu} + \frac{c_{b2}}{c_{b3}}\nabla \tilde{\nu}\cdot\nabla \tilde{\nu} - c_{w1}f_w \left(\frac{\tilde{\nu}}{y}\right)^2
\end{equation}
using the expressions
\begin{alignat*}{5}
\tilde{S} &= S + \frac{\tilde{\nu}}{\kappa^2 y^2} f_{v2} \hspace{1cm} &\bm{\Omega} &= \frac{1}{2}(\nabla \bm{u} - (\nabla \bm{u})^T) \hspace{1cm} & S &= \sqrt{2 \bm{\Omega} : \bm{\Omega}}\\
\chi &= \frac{\tilde{\nu}}{\nu} \hspace{1cm} & f_{v1} &= \frac{\chi^3}{\chi^3 + c_{v1}^3} \hspace{1cm} & f_{v2}  &= 1- \frac{\chi}{1+\chi f_{v1}}\\
f_w &= g \left( \frac{1+c_{w3}^6}{g^6 + c_{w3}^6}\right)^{1/6} \hspace{1cm} & g &=r+c_{w2}(r^6-r) \hspace{1cm} & r &=\frac{\tilde{\nu}}{\tilde{S}\kappa^2 y^2}\\
\intertext{and constants}
c_{b1} &=0.1355 \hspace{1cm} & c_{b2} & = 0.622 \hspace{1cm} & c_{b3} & = 2/3\\
c_{v1} &= 7.1 \hspace{1cm} & \kappa &= 0.41 \\
c_{w1} & = \frac{c_{b1}}{\kappa^2} + \frac{1 + c_{b2}}{c_{b3}} \hspace{1cm} & c_{w2} & = 0.3 & c_{w3} & = 2.
\end{alignat*}
In the non-physical case of $\tilde{\nu}<0$, we set $Q(\bm{u},\tilde{\nu}) = 0$ as well as $\nu_t=0$ according to Crivellini et al.~\cite{Crivellini13}.

\section{Temporal discretization}
\label{sec:temp}
The governing equations~\eqref{eq:conti}--\eqref{eq:mom} are integrated in time as follows. The Navier--Stokes equations are treated using the high-order semi-explicit velocity-correction scheme by Karniadakis et al.~\cite{Karniadakis91}, which is composed of three sub-steps. To avoid non-linear iterations within the Spalart--Allmaras equation, the latter is integrated in time explicitly.

\paragraph{Explicit Spalart--Allmaras step}

The semi-discrete form of the Spalart--Allmaras equation becomes
\begin{multline}
\frac{\gamma_0 \tilde{\nu}^{n+1}-\sum_{i=0}^{J-1}\left(\alpha_i \tilde{\nu}^{n-i}\right)}{\Delta t} = \\
- \sum_{i=0}^{J-1}\beta_i \left(\nabla \cdot \left(\tilde{\bm{\mathcal{F}}}^c\left(\bm{u}^{n-i},\tilde{\nu}^{n-i}\right)-\tilde{\bm{\mathcal{F}}}^{\tilde{\nu}}\left(\tilde{\nu}^{n-i}\right)\right)-Q\left(\bm{u}^{n-i},\tilde{\nu}^{n-i}\right)\right).
\label{eq:sastep}
\end{multline}
with the BDF and EX schemes outlined in Section~\ref{sec:mat}, yielding $\tilde{\nu}^{n+1}$ at the new time step.

\paragraph{Explicit convective step}

The temporal integration of the Navier--Stokes equations follows~\cite{Karniadakis91}. The convective term is integrated explicitly with the same scheme as the Spalart--Allmaras equation, yielding the first intermediate velocity $\hat{\bm{u}}$:
\begin{equation}
\frac{\gamma_0 \hat{\bm{u}}-\sum_{i=0}^{J-1}\left(\alpha_i \bm{u}^{n-i}\right)}{\Delta t} = - \sum_{i=0}^{J-1}\beta_i \nabla \cdot \bm{\mathcal{F}}^c\left(\bm{u}^{n-i}\right)+\bm{f}^{n+1}.
\label{eq:convstep}
\end{equation}

\paragraph{Pressure Poisson equation and projection}

The pressure is subsequently computed by solving a Poisson equation, given as
\begin{equation}
-\nabla^2 p^{n+1} = -\frac{\gamma_0}{\Delta t} \nabla \cdot \hat{\bm{u}}.
\label{eq:poisson}
\end{equation}
For temporal high-order accuracy, consistent boundary conditions for the pressure are essential. They are defined as
\begin{equation}
\nabla p^{n+1} \cdot \bm{n}= -\left(\sum_{i=0}^{J-1}{\beta_i\left(\nabla \cdot \bm{\mathcal{F}}^c\left(\bm{u}_h^{n-i}\right) + \nu\nabla \times \left(\nabla \times \bm{u}^{n-i}\right)\right)-\bm{f}^{n+1}}\right)\cdot \bm{n}
\label{eq:bc_d_pres}
\end{equation}
on $\partial \Omega^D$. Only the solenoidal part of the viscous term is taken into consideration and we make use of the boundary conditions $\bm{g}_{\bm{u}}=\bm{0}$ and $g_{\tilde{\nu}}=0$. With the pressure available, a divergence-free velocity is obtained by
\begin{equation}
\hat{\hat{\bm{u}}}=\hat{\bm{u}}-  \frac{\Delta t}{ \gamma_0} \nabla p^{n+1}
\end{equation}
resulting in the second intermediate velocity $\hat{\hat{\bm{u}}}$.

\paragraph{Implicit viscous step}

Finally, a Helmholtz-like equation is solved for the velocity $\bm{u}^{n+1}$ at time level $t^{n+1}$, given as
\begin{equation}
\frac{\gamma_0}{\Delta t}\left(\bm{u}^{n+1}-\hat{\hat{\bm{u}}}\right) = \nabla \cdot \bm{\mathcal{F}}^{\nu}\left(\bm{u}^{n+1}\right).
\label{eq:visc}
\end{equation}
The system is closed with boundary conditions on the no-slip walls
\begin{align}
\bm{u}^{n+1}&=\bm{0}\text{ \hspace{0.2cm} on } \partial \Omega^D  \text{ \hspace{0.2cm} and } \\
\tilde{\nu}^{n+1}&={0}\text{ \hspace{0.2cm} on } \partial \Omega^D.
\label{eq:bc_vel}
\end{align}

\section{Galerkin formulation}
\label{sec:gal}
Based on the time integration scheme in Appendix~\ref{sec:temp}, weak forms are derived for each sub-step. Besides the standard volume integrals, face integrals are formed on interior boundaries $\partial \Omega_e^{\Gamma}=\partial \Omega_e^-\cap \partial \Omega_e^+$ between two adjacent elements $\Omega_e^-$ and $\Omega_e^+$. The corresponding unit normal vectors $\bm{n}_{\Gamma}^-$ and $\bm{n}_{\Gamma}^+$ point outwards of the respective element with $\bm{n}_{\Gamma}^-=-\bm{n}_{\Gamma}^+$. For the sake of simplicity, the superscripts $(\cdot)^-$ and $(\cdot)^+$ are omitted if possible in the following element-wise variational forms, defining the current element as $\Omega_e^-$ and the neighboring element as $\Omega_e^+$. Jump operators are defined as $[\phi]=\phi^--\phi^+$ and $\llbracket \phi \rrbracket = \phi^- \otimes \bm{n}_{\Gamma}^- + \phi^+ \otimes \bm{n}_{\Gamma}^+$ and an averaging operator as $\{\{\phi\}\}=w^-\phi^-+w^+\phi^+$ with $w^-=w^+=1/2$ if not specified otherwise. At external boundaries, we usually give a suitable definition for $\phi^+$, if not we use $[\phi]=0$ and $\{\{\phi\}\}=\phi^-$ on $\partial \Omega_h^D$. Finally, the $L^2$ inner products presented in the following are abbreviated as usual, e.g. for volume integrals with $(a,b)_{\Omega_e}=\int_{\Omega_e}a b \ d\Omega$ for scalars, $(\bm{a},\bm{b} )_{\Omega_e}=\int_{\Omega_e}\bm{a} \cdot \bm{b} \ d\Omega$ for vectors as well as $(\bm{a},\bm{b} )_{\Omega_e}=\int_{\Omega_e}\bm{a} : \bm{b} \ d\Omega$ for tensors.

The following variational formulations are derived applying a standard procedure by multiplication of the strong forms~\eqref{eq:sastep}--\eqref{eq:bc_vel} with an appropriate weighting function $\mu_h\in \mathcal{V}_h^{\tilde{\nu}}$, $\bm{v}_h\in \mathcal{V}_h^{\bm{u}}$, or $q_h\in \mathcal{V}_h^{p}$ and integration over one element volume. Upon partial integration, suitable fluxes are specified in order to guarantee a stable numerical method.
\paragraph{Explicit Spalart--Allmaras step}
The variational formulation of the Spalart--Allmaras step becomes
\begin{equation}
\begin{split}
&\left(\mu_h,\frac{\gamma_0 \tilde{\nu}_h-\sum_{i=0}^{J-1}\left(\alpha_i \tilde{\nu}_h^{n-i}\right)}{\Delta t}\right)_{\Omega_e} \\
&= -\sum_{i=0}^{J-1}\beta_i \bigg(-\left(\nabla\mu_h, \tilde{\bm{\mathcal{F}}}^c\left(\bm{u}_h^{n-i},\tilde{\nu}_h^{n-i}\right)\right)_{\Omega_e}+\left(\mu_h,\tilde{\bm{\mathcal{F}}}^{c*}\left(\bm{u}_h^{n-i},\tilde{\nu}_h^{n-i}\right)\bm{n}_{\Gamma}\right)_{\partial \Omega_e}\\
&+ \left(\nabla \mu_h,\tilde{\bm{\mathcal{F}}}^{\tilde{\nu}}\left(\tilde{\nu}_h\right)\right)_{\Omega_e} - \frac{1}{2}\left(\tilde{\bm{\mathcal{F}}}^{\tilde{\nu}}\left(\mu_h\right),\llbracket\tilde{\nu}_h\rrbracket\right)_{\partial \Omega_e} - \left(\mu_h,\tilde{\bm{\mathcal{F}}}^{\tilde{\nu}*}\left(\tilde{\nu}_h\right)\right)_{\partial \Omega_e} \\
&-\left(\mu_h,Q\left(\bm{u}_h^{n-i},\tilde{\nu}_h^{n-i}\right)\right)_{\Omega_e}\bigg).
\end{split}
\label{eq:sastepspat}
\end{equation}
Herein, the Lax--Friedrichs numerical flux is employed for the convective term
\begin{equation}
    \tilde{\bm{\mathcal{F}}}^{c*}\left(\bm{u}_h^{n-i},\tilde{\nu}_h^{n-i}\right)=
                  \{\{\tilde{\bm{\mathcal{F}}}^{c}\left(\bm{u}_h^{n-i},\tilde{\nu}_h^{n-i}\right)\}\}+\tilde{\Lambda}/2\llbracket \tilde{\nu}_h^{n-i}\rrbracket,
\end{equation}
where $\tilde{\Lambda} = \max(\tilde{\lambda}^-,\tilde{\lambda}^+)$
represents the largest eigenvalue of the flux Jacobian across the element interface with
\begin{equation}
\begin{array}{ll}
\tilde{\lambda}^-=\max_j\left|\tilde{\lambda}_j\left(\frac{\partial \tilde{\bm{\mathcal{F}}}\left(\bm{u}_h^{-,n-i},\tilde{\nu}\right)\cdot \bm{n}_{\Gamma}}{\partial \tilde{\nu}}\right)\right| = |\bm{u}_h^{-,n-i} \cdot \bm{n}_{\Gamma}| \text{ and}\\
\tilde{\lambda}^+=\max_j\left|\tilde{\lambda}_j\left(\frac{\partial \tilde{\bm{\mathcal{F}}}\left(\bm{u}_h^{+,n-i},\tilde{\nu}\right)\cdot \bm{n}_{\Gamma}}{\partial \tilde{\nu}}\right)\right| = |\bm{u}_h^{+,n-i} \cdot \bm{n}_{\Gamma}|.
\end{array}
\end{equation}
The diffusive term is discretized by the symmetric interior penalty method~\cite{Arnold82} using harmonic weighting~\cite{Burman12} of the discontinuous diffusivity
\begin{equation}
w^-=\frac{\nu+\tilde{\nu}_h^+}{2\nu+\tilde{\nu}_h^-+\tilde{\nu}_h^+} \hspace{1cm}
w^+=\frac{\nu+\tilde{\nu}_h^-}{2\nu+\tilde{\nu}_h^-+\tilde{\nu}_h^+},
\label{eq:harmweightssa}
\end{equation}
which renders the formulation stable for significant discontinuities in $\tilde{\nu}_h$ as well. The diffusive flux then becomes 
\begin{equation}
    \tilde{\bm{\mathcal{F}}}^{\tilde{\nu}*}\left(\tilde{\nu}_h^{n-i}\right)=
                  \{\{\tilde{\bm{\mathcal{F}}}^{\tilde{\nu}}\left(\tilde{\nu}_h^{n-i}\right)\}\} -\tau_{\mathrm{IP}} \frac{2\left(\nu+\tilde{\nu}_h^-\right)\left(\nu+\tilde{\nu}_h^+\right)}{c_{b3}\left(2\nu+\tilde{\nu}_h^-+\tilde{\nu}_h^+\right)} \llbracket \tilde{\nu}_h^{n-i}\rrbracket.
\end{equation}
The interior penalty parameter definition by~\cite{Hillewaert13} is adopted, reading
\begin{equation}
\tau_{\mathrm{IP}} = \left\{
\begin{array}{ll}
\max\left(\tau_{\mathrm{IP},e}^-,\tau_{\mathrm{IP},e}^+\right) &\text{ \hspace{0.2cm}  on } \partial \Omega_e^{\Gamma} \text{ and}\\
\tau_{\mathrm{IP},e}^- &\text{ \hspace{0.2cm}  on } \partial \Omega_e^{D}
\end{array} \right.
\label{eq:tau}
\end{equation}
with
\begin{equation}
\tau_{\mathrm{IP},e}=(k+1)^2\frac{A\left(\partial \Omega_e^{\Gamma}\right)/2+A\left(\partial \Omega_e^{D}\right)}{V\left(\Omega_e\right)}
\label{eq:taue}
\end{equation}
including face area $A$ and cell volume $V$.
On external boundaries, we define $\tilde{\nu}_h^+=-\tilde{\nu}_h^-$, $\nabla \tilde{\nu}_h^+=\nabla\tilde{\nu}_h^-$ and $\bm{u}_h^+=-\bm{u}_h^-$, which are in agreement with no-slip Dirichlet boundary conditions.

\paragraph{Explicit convective step}

The weak form of the convective step in the Navier--Stokes equation is very similar to the convective term present in the Spalart--Allmaras equation and reads
\begin{multline}
\left(\bm{v}_h,\frac{\gamma_0 \hat{\bm{u}}_h-\sum_{i=0}^{J-1}\left(\alpha_i \bm{u}_h^{n-i}\right)}{\Delta t}\right)_{\Omega_e} \\
= -\sum_{i=0}^{J-1}\beta_i \left(-\left(\nabla\bm{v}_h, \bm{\mathcal{F}}^c\left(\bm{u}_h^{n-i}\right)\right)_{\Omega_e}+\left(\bm{v}_h,\bm{\mathcal{F}}^{c*}\left(\bm{u}_h^{n-i}\right)\bm{n}_{\Gamma}\right)_{\partial \Omega_e}\right)+\left(\bm{v}_h,\bm{f}_h^{n+1}\right)_{\Omega_e}
\label{eq:convstepspat}
\end{multline}
with $\hat{\bm{u}}_h\in\mathcal{V}_h^{\bm{u}}$, with again the Lax--Friedrichs numerical flux
\begin{equation}
    \bm{\mathcal{F}}^{c*}\left(\bm{u}_h^{n-i}\right)=
                  \{\{\bm{\mathcal{F}}^{c}\left(\bm{u}_h^{n-i}\right)\}\}+\Lambda/2\llbracket \bm{u}_h^{n-i}\rrbracket,
\end{equation}
and with $\Lambda = \max(\lambda^-,\lambda^+)$. The maximum eigenvalue of the flux Jacobian is given by
\begin{equation}
\begin{array}{ll}
\lambda^-=\max_j\left|\lambda_j\left(\frac{\partial \bm{\mathcal{F}}\left(\bm{u}\right)\cdot \bm{n}_{\Gamma}}{\partial \bm{u}}\big|_{\bm{u}_h^{-,n-i}}\right)\right| = 2|\bm{u}_h^{-,n-i} \cdot \bm{n}_{\Gamma}| \text{ and}\\
\lambda^+=\max_j\left|\lambda_j\left(\frac{\partial \bm{\mathcal{F}}\left(\bm{u}\right)\cdot \bm{n}_{\Gamma}}{\partial \bm{u}}\big|_{\bm{u}_h^{+,n-i}}\right)\right| = 2|\bm{u}_h^{+,n-i} \cdot \bm{n}_{\Gamma}|.
\end{array}
\end{equation}
No-slip Dirichlet boundary conditions are imposed by computing the external velocity as $\bm{u}_h^+ = -\bm{u}_h^-$.

\paragraph{Poisson equation and projection}
The second derivatives present in the pressure Poisson equation are again discretized using the symmetric interior penalty method~\cite{Arnold82}. We further adopt the suggestions presented in~\cite{Krank16b} regarding the partial integration of the right-hand side of the Poisson equation.
We get
\begin{multline}
\left(\nabla q_h,\nabla p_h^{n+1}\right)_{\Omega_e} - \frac{1}{2}\left(\nabla q_h,\llbracket p_h^{n+1}\rrbracket\bm{n}_{\Gamma} \right)_{\partial \Omega_e} - \left(q_h,\bm{\mathcal{P}}^*\cdot \bm{n}_{\Gamma}\right)_{\partial \Omega_e}\\
=\left(\nabla q_h,\frac{\gamma_0}{\Delta t} \hat{\bm{u}}_h\right)_{\Omega_e} - \left(q_h,\frac{\gamma_0}{\Delta t}\{\{\hat{\bm{u}}_h\}\}\cdot \bm{n}_{\Gamma}\right)_{\partial \Omega_e}
\label{eq:poissonspat}
\end{multline}
with the flux function including the high-order boundary conditions~\eqref{eq:bc_d_pres}
\begin{equation}
    \bm{\mathcal{P}}^*=\left\{
                \begin{array}{ll}
                  \{\{\nabla p_h^{n+1}\}\}-\tau_{\mathrm{IP}}\llbracket p_h^{n+1} \rrbracket  &\text{ \hspace{0.1cm}  on }\partial \Omega_e^{\Gamma} \text{ and}\\
                  -\left(\sum_{i=0}^{J-1}{\beta_i\left(\nabla \cdot \bm{\mathcal{F}}^c\left(\bm{u}_h^{n-i}\right)+\nu\nabla \times \bm{\omega}_h^{n-i}\right)-\bm{f}^{n+1}}\right) &\text{ \hspace{0.1cm}  on } \partial \Omega_e^{D}.
                \end{array}
              \right.
              \label{eq:ipflux}
\end{equation}
The same interior penalty parameter definition as in~\eqref{eq:tau} is adopted for $\tau_{\mathrm{IP}}$ and the vorticity $\bm{\omega}_h$ contained in the latter is pre-computed via $L^2$ projection as in~\cite{Krank16b} such that the second derivatives do not have to be computed directly. On $\partial \Omega_e^{D}$ the exterior pressure is further set to $p_h^+=p_h^-$.

The subsequent projection step is defined as in~\cite{Krank16b} including a div-div penalty term and a partially integrated right-hand side, yielding
\begin{multline}
\left(\bm{v}_h,\hat{\hat{\bm{u}}}_h\right)_{\Omega_e} + \left(\nabla \cdot \bm{v}_h, \tau_{\mathrm{D}}\nabla \cdot \hat{\hat{\bm{u}}}_h\right)_{\Omega_e} \\
= \left(\bm{v}_h,\hat{\bm{u}}_h\right)_{\Omega_e} + \left(\nabla \cdot \bm{v}_h,\frac{\Delta t}{\gamma_0} p_h^{n+1}\right)_{\Omega_e} - \left(\bm{v}_h,\frac{\Delta t}{\gamma_0} \{\{p_h^{n+1}\}\} \bm{n}_{\Gamma}\right)_{\partial \Omega_e},
\label{eq:projectionspat}
\end{multline}
where $\hat{\hat{\bm{u}}}_h\in\mathcal{V}_h^{\bm{u}}$ and the penalty parameter $\tau_{\mathrm{D}}$ is defined according to~\cite{Krank16b}
\begin{equation}
\tau_{\mathrm{D}}=\frac{ 10\Vert \left<{\bm{u}}_h^n\right>\Vert h \Delta t}{\mathrm{CFL}},
\label{eq:tauD}
\end{equation}
with the element-wise mean velocity $\left<\bm{u}_h\right>$, $h=V^{1/d}$ based on the element volume and an amplification factor of $10$ for stability in all configurations.
\paragraph{Implicit viscous step}

The viscous term is discretized by the non-symmetric interior penalty method~\cite{Riviere99}, as detailed in Section~\ref{sec:mat}, since the enrichment relies on a non-polynomial function space and the definition of $\tau_{\mathrm{IP}}$ in~\eqref{eq:tau} would not be appropriate. We get
\begin{multline}
\left(\bm{v}_h,\frac{\gamma_0}{\Delta t}\bm{u}_h^{n+1}\right)_{\Omega_e} + \left(\bm{\epsilon}\left(\bm{v}_h\right),\bm{\mathcal{F}}^{\nu}\left(\bm{u}_h^{n+1}\right)\right)_{\Omega_e} - \frac{s}{2}\left(\bm{\mathcal{F}}^{\nu}\left(\bm{v}_h\right),\llbracket\bm{u}_h^{n+1}\rrbracket\right)_{\partial \Omega_e}
\\ - \left(\bm{v}_h,\bm{\mathcal{F}}^{\nu*}\left(\bm{u}_h^{n+1}\right)\cdot \bm{n}_{\Gamma}\right)_{\partial \Omega_e} = 
 \left(\bm{v}_h,\frac{\gamma_0}{\Delta t}\hat{\hat{\bm{u}}}_h\right)_{\Omega_e}.
\end{multline}
and choose $s=-1$. Again we deal with possible discontinuities present in the viscosity using harmonic weighting according to~\eqref{eq:harmweightssa} by replacing $\tilde{\nu}_h$ with $\nu_t$. The interior penalty flux is defined as
\begin{equation}
    \bm{\mathcal{F}}^{\nu*}(\bm{u}_h^{n+1})=
                  \{\{\bm{\mathcal{F}}^{\nu}(\bm{u}_h^{n+1})\}\} -\tau_{\mathrm{IP}} \frac{2(\nu+\nu_t^+)(\nu+\nu_t^-)}{2\nu + \nu_t^-+\nu_t^+} \llbracket \bm{u}_h^{n+1}\rrbracket
\end{equation}
and we take the same value for $\tau_{\mathrm{IP}}$ as previously~\eqref{eq:tau}, as this penalization yields good results in our computational experiments. A drawback of this non-symmetric formulation is that we may only obtain convergence rates of the baseline solver of order $k$ in the $L^2$ norm~\cite{Hartmann07}, in contrast to $k+1$ of the original version of the solver employing the symmetric interior penalty method in~\cite{Krank16b}.

Our preliminary investigations have shown that the weak boundary conditions suggested in~\cite{Krank16b} result in pronounced discontinuities with the present enrichment approach due to the large velocity gradients at the wall. We therefore employ strong velocity Dirichlet boundary conditions on the viscous terms, prescribing $\bm{u}_h^+ = \bm{u}_h^- = \bm{0}$ with $\nabla \bm{u}_h^+=\nabla \bm{u}_h^-$ on the face terms whereas no changes are made in the mass term.

\section{Time stepping algorithm}
\label{sec:dt_algo}
The aim of this adaptive time stepping algorithm is to make optimal use of the allowable time step size with regard to the CFL condition as well as the diffusion number.
The CFL condition is in the present work defined as
\begin{equation}
\frac{\mathrm{CFL}}{k^{1.5}} = \max_j\left|(\bm{J}^{-T}\bm{u}_h^n)_j\right| \Delta t^{\mathrm{NS}}
\end{equation}
where the transposed inverse of the Jacobian is employed to transform $\bm{u}_h^n$ into the parameter space of each element. Here, the largest absolute vector component of $\bm{J}^{-T}\bm{u}_h^n$ as velocity-to-length ratio has performed best in our investigations regarding curved boundaries and anisotropic meshes. The exponent of the polynomial degree of $1.5$ has been determined experimentally as giving the tightest fit for $k\leq8$ and is lower than the theoretical value of $2$ frequently used (see, e.g., \cite{Krank16b}).
The time step size $\Delta t^{\mathrm{NS}}$ resulting from a constant CFL number is computed in each step and applied for the time-advancement of the Navier--Stokes equations.

The Spalart--Allmaras equation is additionally subject to the diffusion number $D$, which is given as
\begin{equation}
\frac{D}{k^3} = \frac{(\nu + \tilde{\nu}_h^n)\Delta t^{\mathrm{SA}}}{c_{b3}h^2}
\end{equation}
with the cell-wise shortest edge length $h$ and again an experimentally determined exponent of $k$ chosen as 3. As the Spalart--Allmaras equation is much cheaper to evaluate compared to the Navier--Stokes equation, a sub-cycling algorithm is employed in case the latter condition restricts the time-step size, keeping the convective velocity constant during the sub-cycles. The number of sub-cycles is given by $N^{\mathrm{SA}}=\max(1,\lceil \Delta t^{\mathrm{NS}}/\Delta t^{\mathrm{SA}}\rceil)$ and defines the final time step to be used for the Spalart--Allmaras equation within each Navier--Stokes time step to $\Delta t^{\mathrm{SA}} = \Delta t^{\mathrm{NS}}/N^{\mathrm{SA}}$. The time integrator constants $\alpha_i$, $\beta_i$ and $\gamma_0$ for non-equally spaced time intervals may be determined in a straight-forward manner, for example via Taylor expansion, and are re-calculated when necessary. Sub-cycling counts in the simulations of the present paper are in the interval $N^{\mathrm{SA}}=[1,9]$. If significantly larger $N^{\mathrm{SA}}$ occur, one may consider to use an implicit implementation of the Spalart--Allmaras equation.

\acks
The research presented in this article was partially funded by the German Research Foundation (DFG) under the project ``High-order discontinuous Galerkin for the EXA-scale'' (ExaDG) within the priority program ``Software for Exascale Computing'' (SPPEXA), grant agreement no. KR4661/2-1 and WA1521/18-1. The authors would like to thank Stefan Legat for preparations towards the periodic hill test case as well as the time adaptation scheme.

\bibliographystyle{wileyj}
\bibliography{dg_rans_paper_v1}

\begin{thebibliography}{10}
\providecommand{\url}[1]{\texttt{#1}}
\providecommand{\urlprefix}{URL }
\expandafter\ifx\csname urlstyle\endcsname\relax
  \providecommand{\doi}[1]{doi:\discretionary{}{}{}#1}\else
  \providecommand{\doi}{doi:\discretionary{}{}{}\begingroup
  \urlstyle{rm}\Url}\fi

\bibitem{Launder74}
Launder BE, Spalding DB. The numerical computation of turbulent flows.
  \emph{Computer Methods in Applied Mechanics and Engineering}  1974;
  \textbf{3}(2):269 -- 289, \doi{10.1016/0045-7825(74)90029-2}.

\bibitem{Durbin09}
Durbin PA. Limiters and wall treatments in applied turbulence modeling.
  \emph{Fluid Dynamics Research}  2009; \textbf{41}(1):012\,203.

\bibitem{Kalitzin05}
Kalitzin G, Medic G, Iaccarino G, Durbin P. Near-wall behavior of {RANS}
  turbulence models and implications for wall functions. \emph{Journal of
  Computational Physics}  2005; \textbf{204}(1):265 -- 291,
  \doi{10.1016/j.jcp.2004.10.018}.

\bibitem{Vieser02}
Vieser W, Esch T, Menter F. Heat transfer predictions using advanced
  two-equation turbulence models. \emph{CFX Validation Report, Report No.
  CFX-VAL}  2002; \textbf{10}:0602.

\bibitem{Knopp06}
Knopp T, Alrutz T, Schwamborn D. A grid and flow adaptive wall-function method
  for {RANS} turbulence modelling. \emph{Journal of Computational Physics}
  2006; \textbf{220}(1):19 -- 40, \doi{10.1016/j.jcp.2006.05.003}.

\bibitem{Popovac07}
Popovac M, Hanjalic K. Compound wall treatment for {RANS} computation of
  complex turbulent flows and heat transfer. \emph{Flow, Turbulence and
  Combustion}  2007; \textbf{78}(2):177--202, \doi{10.1007/s10494-006-9067-x}.

\bibitem{Melenk96}
Melenk J, Babu\v{s}ka I. The partition of unity finite element method: {B}asic
  theory and applications. \emph{Computer Methods in Applied Mechanics and
  Engineering}  1996; \textbf{139}(1–4):289 -- 314,
  \doi{10.1016/S0045-7825(96)01087-0}.

\bibitem{Belytschko99}
Belytschko T, Black T. Elastic crack growth in finite elements with minimal
  remeshing. \emph{International Journal for Numerical Methods in Engineering}
  1999; \textbf{45}(5):601--620,
  \doi{10.1002/(SICI)1097-0207(19990620)45:5<601::AID-NME598>3.0.CO;2-S}.

\bibitem{Fries10}
Fries TP, Belytschko T. The extended/generalized finite element method: an
  overview of the method and its applications. \emph{International Journal for
  Numerical Methods in Engineering}  2010; \textbf{84}(3):253--304,
  \doi{10.1002/nme.2914}.

\bibitem{Turner10}
Turner D, Noble D. Analytic enrichment of finite element formulations for
  capturing boundary layer behavior. AIAA paper no. 2010-4996, 2010,
  \doi{10.2514/6.2010-4996}.

\bibitem{Chen14}
Chen L, Edeling WN, Hulshoff SJ. {POD} enriched boundary models and their
  optimal stabilisation. \emph{International Journal for Numerical Methods in
  Fluids}  2014; \textbf{77}(2):92--107, \doi{10.1002/fld.3977}.

\bibitem{Farhat01}
Farhat C, Harari I, Franca LP. The discontinuous enrichment method.
  \emph{Computer Methods in Applied Mechanics and Engineering}  2001;
  \textbf{190}(48):6455 -- 6479, \doi{10.1016/S0045-7825(01)00232-8}.

\bibitem{Kalashnikova09}
Kalashnikova I, Farhat C, Tezaur R. A discontinuous enrichment method for the
  finite element solution of high {P}{\'e}clet advection--diffusion problems.
  \emph{Finite Elements in Analysis and Design}  2009; \textbf{45}(4):238--250,
  \doi{10.1016/j.finel.2008.10.009}.

\bibitem{Krank16}
Krank B, Wall WA. A new approach to wall modeling in {LES} of incompressible
  flow via function enrichment. \emph{Journal of Computational Physics}  2016;
  \textbf{316}:94 -- 116, \doi{10.1016/j.jcp.2016.04.001}.

\bibitem{Krank16b}
Krank B, Fehn N, Wall WA, Kronbichler M. A high-order semi-explicit
  discontinuous {G}alerkin solver for 3{D} incompressible flow with application
  to {DNS} and {LES} of turbulent channel flow. \emph{arXiv preprint
  arXiv:1607.01323}  2016; .

\bibitem{Spalart94}
Spalart PR, Allmaras SR. A one-equation turbulence model for aerodynamic flows.
  \emph{La Recherche A\'erospatiale}  1994; \textbf{1}:5--21.

\bibitem{Moro11}
Moro D, Nguyen N, Peraire J. Navier--{S}tokes solution using hybridizable
  discontinuous {G}alerkin methods. AIAA paper no. 2011-3407, 2011,
  \doi{10.2514/6.2011-3407}.

\bibitem{Burgess12}
Burgess NK, Mavriplis DJ. Robust computation of turbulent flows using a
  discontinuous {G}alerkin method. AIAA paper no. 2012-457, 2012,
  \doi{10.2514/6.2012-457}.

\bibitem{Crivellini13}
Crivellini A, D'Alessandro V, Bassi F. A {S}palart–-{A}llmaras turbulence
  model implementation in a discontinuous {G}alerkin solver for incompressible
  flows. \emph{Journal of Computational Physics}  2013; \textbf{241}:388 --
  415, \doi{10.1016/j.jcp.2012.12.038}.

\bibitem{Darmofal13}
Darmofal DL, Allmaras SR, Yano M, Kudo J. An adaptive, higher-order
  discontinuous {G}alerkin finite element method for aerodynamics. AIAA paper
  no. 2013-2871, 2013, \doi{10.2514/6.2013-2871}.

\bibitem{ZhenHua16}
ZhenHua J, Chao Y, Jian Y, Feng Q, Wu Y. A {S}palart--{A}llmaras turbulence
  model implementation for high-order discontinuous {G}alerkin solution of the
  {R}eynolds-averaged {N}avier-{S}tokes equations. \emph{Flow, Turbulence and
  Combustion}  2016; \textbf{96}(3):623--638, \doi{10.1007/s10494-015-9656-7}.

\bibitem{Dean76}
Dean RB. A single formula for the complete velocity profile in a turbulent
  boundary layer. \emph{Journal of Fluids Engineering}  1976;
  \textbf{98}(4):723--726, \doi{10.1115/1.3448467}.

\bibitem{Spalding61}
Spalding DB. A single formula for the ``law of the wall''. \emph{Journal of
  Applied Mechanics}  1961; \textbf{28}:455--458, \doi{10.1115/1.3641728}.

\bibitem{Dean78}
Dean RB. Reynolds number dependence of skin friction and other bulk flow
  variables in two-dimensional rectangular duct flow. \emph{Journal of Fluids
  Engineering}  1978; \textbf{100}(2):215--223, \doi{10.1115/1.3448633}.

\bibitem{Karniadakis91}
Karniadakis GE, Israeli M, Orszag SA. High-order splitting methods for the
  incompressible {N}avier--{S}tokes equations. \emph{Journal of Computational
  Physics}  1991; \textbf{97}(2):414 -- 443,
  \doi{10.1016/0021-9991(91)90007-8}.

\bibitem{Arnold82}
Arnold DN. An interior penalty finite element method with discontinuous
  elements. \emph{SIAM Journal on Numerical Analysis}  1982;
  \textbf{19}(4):742--760, \doi{10.1137/0719052}.

\bibitem{Riviere99}
Rivi{\`e}re B, Wheeler MF, Girault V. Improved energy estimates for interior
  penalty, constrained and discontinuous {G}alerkin methods for elliptic
  problems. {P}art {I}. \emph{Computational Geosciences}  1999;
  \textbf{3}(3):337--360, \doi{10.1023/A:1011591328604}.

\bibitem{Burman12}
Burman E, Zunino P. Numerical approximation of large contrast problems with the
  unfitted {N}itsche method. \emph{Frontiers in Numerical Analysis - Durham
  2010}, Blowey J, Jensen M (eds.), Springer, 2012; 227--282,
  \doi{10.1007/978-3-642-23914-4_4}.

\bibitem{Kronbichler12}
Kronbichler M, Kormann K. A generic interface for parallel cell-based finite
  element operator application. \emph{Computers \& Fluids}  2012;
  \textbf{63}:135--147, \doi{10.1016/j.compfluid.2012.04.012}.

\bibitem{Bangerth16}
Bangerth W, Davydov D, Heister T, Heltai L, Kanschat G, Kronbichler M, Maier M,
  Turcksin B, Wells D. The deal.{II} library, version 8.4.0. \emph{Journal of
  Numerical Mathematics}  2016; \textbf{24}(3):135--141,
  \doi{10.1515/jnma-2016-1045}.

\bibitem{Davydov16}
Davydov D, Gerasimov T, Pelteret JP, Steinmann P. On the $h$ -adaptive {PUM}
  and the $hp$ -adaptive {FEM} approaches applied to {PDEs} in quantum
  mechanics  2016; Submitted to Journal of Computational Physics.

\bibitem{Moser99}
Moser RD, Kim J, Mansour NN. Direct numerical simulation of turbulent channel
  flow up to {Re}$_{\tau}$=590. \emph{Physics of Fluids}  1999;
  \textbf{11}(4):943--945, \doi{10.1063/1.869966}.

\bibitem{Alamo03}
Del~{\'A}lamo JC, Jim{\'e}nez J. Spectra of the very large anisotropic scales
  in turbulent channels. \emph{Physics of Fluids}  2003;
  \textbf{15}(6):L41--L44, \doi{10.1063/1.1570830}.

\bibitem{Hoyas06}
Hoyas S, Jim{\'e}nez J. Scaling of the velocity fluctuations in turbulent
  channels up to {Re}$_{\tau}$=2003. \emph{Physics of Fluids}  2006;
  \textbf{18}(1):011702, \doi{10.1063/1.2162185}.

\bibitem{Lee15}
Lee M, Moser RD. Direct numerical simulation of turbulent channel flow up to
  $\mathit{Re}_{{\it\tau}}\approx 5200$. \emph{Journal of Fluid Mechanics}
  2015; \textbf{774}:395–415, \doi{10.1017/jfm.2015.268}.

\bibitem{Frohlich05}
Fr{\"o}hlich J, Mellen CP, Rodi W, Temmerman L, Leschziner MA. Highly resolved
  large-eddy simulation of separated flow in a channel with streamwise periodic
  constrictions. \emph{Journal of Fluid Mechanics}  2005; \textbf{526}:19--66,
  \doi{10.1017/S0022112004002812}.

\bibitem{Rapp11}
Rapp C, Manhart M. Flow over periodic hills: an experimental study.
  \emph{Experiments in Fluids}  2011; \textbf{51}(1):247--269,
  \doi{10.1007/s00348-011-1045-y}.

\bibitem{Jakirlic15}
Jakirli\'c S, Maduta R. Extending the bounds of ‘steady’ {RANS} closures:
  Toward an instability-sensitive {R}eynolds stress model. \emph{International
  Journal of Heat and Fluid Flow}  2015; \textbf{51}:175 -- 194,
  \doi{10.1016/j.ijheatfluidflow.2014.09.003}.

\bibitem{RappERCOFTAC}
Rapp C, Breuer M, Manhart M, Peller N. 2{D} {P}eriodic {H}ill {F}low 2010.
  \urlprefix\url{http://qnet-ercoftac.cfms.org.uk/}, last accessed 17 October
  2016.

\bibitem{Jakirlic10}
Jakirli\'c S. Periodic flow over a 2-{D} hill {C}ross-plots ${R}e_{H}=10600$
  2010.
  \urlprefix\url{http://cfd.mace.manchester.ac.uk/twiki/bin/view/ATAAC/TestCase001PeriodicHill},
  {S}lides of presentation held at Chalmers University in Gothenburg, Sweden,
  in April 2010, last accessed 17 October 2016.

\bibitem{Spalart97}
Spalart PR, Jou W, Strelets M, Allmaras SR. {Comments on the feasibility of LES
  for wings, and on a hybrid RANS/LES approach}. \emph{Advances in DNS/LES},
  Liu C, Liu Z (eds.). Greyden Press Columbus, OH, 1997.

\bibitem{Hillewaert13}
Hillewaert K. Development of the discontinuous {G}alerkin method for
  high-resolution, large scale {CFD} and acoustics in industrial geometries.
  Ph{D} {T}hesis, Univ. de Louvain 2013.

\bibitem{Hartmann07}
Hartmann R. Adjoint consistency analysis of discontinuous {G}alerkin
  discretizations. \emph{SIAM Journal on Numerical Analysis}  2007;
  \textbf{45}(6):2671--2696, \doi{10.1137/060665117}.

\end{thebibliography}
\end{document}